\documentclass[11pt]{article}
\usepackage[affil-it]{authblk}
\usepackage{array,multirow,graphicx}
\usepackage{color}
\usepackage{times}
\usepackage{amsmath}
\usepackage{amsfonts}
\usepackage{enumerate}
\usepackage{rotating}
\usepackage{url}
\usepackage{natbib}

\def\({\left(}
\def\){\right)}
\def\[{\left[}
\def\]{\right]}

\newcommand{\bepsilon}{\boldsymbol{\epsilon}}

\newcommand{\bPhi}{\boldsymbol{\Phi}}
\newcommand{\0}{\mathbf{0}}

\newcommand{\blue}{\textcolor{blue}}

\def\be{\begin{eqnarray}}
\def\ee{\end{eqnarray}}
\def\bq{\begin{equation}}
\def\eq{\end{equation}}
\def\bse{\begin{eqnarray*}}
\def\ese{\end{eqnarray*}}

\def\th{\mathrm{th}}

\def\bX{\mathbf{X}}
\def\bY{\mathbf{Y}}
\def\by{\mathbf{y}}

\def\bdPhi{\boldsymbol{\Phi}}

\usepackage[normalem]{ulem}

\newcommand\startsupplement{%
    \makeatletter 
       \setcounter{table}{0}
       \setcounter{figure}{0}
       \renewcommand{\thefigure}{S\arabic{figure}}
	   \renewcommand{\thetable}{S\arabic{table}}
    \makeatother}

\begin{document}

\title{Bayesian functional graphical models}
\author{Lin Zhang$^{1*}$, Veera Baladandayuthapani$^2$, 
Quinton Neville$^1$, Karina Quevedo $^3$, Jeffrey S. Morris $^4$}
\affil{$^1$Division of Biostatistics, University of Minnesota, Minneapolis, MN, U.S.A. \\
$^2$Department of Biostatistics, University of Michigan, Ann Arbor, MI, U.S.A. \\
$^3$Department of Psychiatry and Behavioral Sciences, University of Minnesota, Minneapolis, MN, U.S.A. \\
$^4$ Department of Biostatistics, Epidemiology and Informatics, Perelman School of Medicine, University of Pennsylvania, Philadelphia, PA, U.S.A. \\
$^*$ To whom correspondence should be addressed (\texttt{zhan4800@umn.edu}) }
\date{}

\maketitle

\begin{abstract}
We develop a Bayesian  graphical modeling framework for  functional data for correlated multivariate random variables observed over a continuous domain. Our method leads to graphical Markov models for functional data which allows the graphs to vary over the functional domain.  The model involves estimation of graphical models that evolve functionally in a nonparametric fashion while accounting for within-functional correlations and borrowing strength across functional positions so contiguous locations are encouraged but not forced to have similar graph structure and edge strength.  We utilize a strategy that combines nonparametric basis function modeling with modified Bayesian graphical regularization techniques, which induces a new class of hypoexponential normal scale mixture distributions that not only leads to adaptively shrunken estimators of the conditional cross-covariance but also facilitates a thorough theoretical investigation of the shrinkage properties.  Our approach scales up to large functional datasets collected on a fine grid. We show through simulations and real data analysis that the Bayesian functional graphical model can efficiently reconstruct the functionally-evolving graphical models by accounting for within-function correlations.  \\
\blue{keywords}: Bayesian graphical modeling; functional data; functional graphical model; normal-hypoexponential; shrinkage priors.
\end{abstract}

\section{Introduction}
Development of high-throughput technologies avails large amounts of high-dimensional data that involve complex correlation structures. Graphical models provide a general framework for inferring the conditional dependence structures among a set of random variables. For a $p$-dimensional random vector $\bY=(Y_1,\dots,Y_p)$, a graphical model $G=(V,E)$ consists of a vertex set $V$ including the $p$ random variables and an edge set $E$ respecting the conditional dependencies among the variables. In Gaussian graphical models, the problem of learning a graph is equivalent to estimation of the covariance matrix or its inverse. Common approaches include Bayesian hierarchical models using the hyper-inverse Wishart (HIW) priors \citep{lauritzen1996, armstrong2009}, and the graphical lasso method \citep{Yuan2007glasso, Friedman2007glasso} and its Bayesian extensions \citep{wang2012glasso, Baladandayuthapani2014, Zhang2016covdecomp}.  These methods are used to estimate a static, population level graphs across subjects assuming a stable state. In this article, we focus on functional graphical modeling, in which the observed datum $\bY_i(t) = [Y_{i1}(t), \ldots, Y_{ip}(t)]$ are functions observed on a domain $\mathcal{T}$ and interest lies in the graphical model that varies along the functional domain $G(t) = (V,E(t))$.

The problem of functional graphical modeling arises from multivariate functional data analysis with repeated measurements of multiple variables at a series of distinct time points, examples of which include time-series gene expression data and longitudinal neuroimaging data. A large proportion of related research has mostly focused on time-varying graphical models. \citet{cribben2012, robinson2015} proposed piece-wise constant graphical models that partition the time course into temporal intervals with constant network in each interval. The Hidden Markov Models, on the other hand, assume presence of latent states with associated graphs and estimate dynamic changes in state and corresponding graphical models \citep{rack2012, warnick2018}.  Other methods assume smooth changes of network structures over a functional domain, which can be broadly classified to sliding windows approaches \citep{kucyi2014, elton2015}, kernel-based nonparametric methods \citep{zhou2010, kolar2011, gibberd2017}, and fused lasso type methods \citep{yang2015fglasso, hallac2017, yang&peng2020}. Most of these methods are for single-subject data analysis and are not straightforward to obtain common network changes of groups of subjects. In addition, these methods assume independence of data across locations, which is usually not true for functional data, and utilize only local data for graph estimation at a certain location. \citet{qiao2019} and \citet{zhu2016} developed methods utilizing functional principal component or orthogonal basis function representation combined with the group lasso method or hyper-inverse-Wishart prior, which account for within-functional correlations and infer common network structure in multi-subject analysis. However, both methods assumes a constant network structure over the function domain among the variables.

In this paper, we propose a Bayesian functional graphical model for inference of common network evolutions among a group of subjects which accounts for within-functional correlations that are commonly present in multivariate time-series data and incorporates both local and global information in graph estimation at each location. The Bayesian functional graphical model induces locally adaptive shrinkage on conditional cross-covariance that leads to functionally-evolving network estimate by shrinking the conditional covariance of basis function coefficients. Compared to existing methods, our model (1) accounts for correlations across different locations that are ignored by time-varying graphical models; (2) pools information from both local and global regions in graphical model estimation at each location; (3) allows for flexible nonparametric representation of the functional curves rather than pre-specified parametric functional forms; and (4) can scale up to large functional datasets collected on a fine grid.   Most importantly, we show theoretically that the hierarchical model specified in the dual basis space induces a normal scale mixture prior distribution in the data space with locally adaptive shrinkage of conditional cross-covariance matrices at each location. We utilize a variant of the Bayesian graphical lasso method in the basis space that induces a scale-mixture prior, the hypoexponential distribution, in the data space, of which the adaptive shrinkage properties on the conditional cross-covariance  were examined theoretically. Our simulation and real data analyses show that compared to competitive methods, our Bayesian functional graphical model has higher power in detecting functional changes of connections by accounting for both short- and long-range within-functional correlations.

The outline for the rest of the paper is as follows. In Section \ref{sec:model}, we present our Bayesian functional graphical model for multivariate functional data analysis. We discuss the posterior inference methods in Section \ref{sec:postinf}. We present the results of a simulation study to examine the performance of our method in Section \ref{sec:simu}, and apply the method to a task-based functional magnetic resonance imaging (fMRI) dataset in Section \ref{sec:app}. The paper finally concludes with a discussion in Section \ref{sec:diss}. 

\section{The functional graphical model} \label{sec:model}

Suppose that $y_{ij}(t)$ is a functional curve observed for sample $i$ $(i=1,\ldots,N)$ and variable $j$ $(j=1,\ldots,p)$ on a common interval $\mathcal{T}$. We assume that the $p \times 1$ vector of observed functions, $\by_i(t) = [y_{i1}(t), \ldots, y_{ip}(t)]$, are realizations of the set of the random functional variables $\{Y_j(t)\}_{j=1,\ldots,p}$ following a multivariate Gaussian random process. Following \citet{qiao2019}, we define the conditional cross-covariance function for any two variables as, 
\bse
C_{j\ell}(t,t^\prime) = Cov\{Y_j(t),Y_\ell(t^\prime)|Y_h(s), \; h\neq j,\ell, \; \forall s \in \mathcal{T} \}
\ese
which gives the partial covariance between $Y_j(t)$ and $Y_\ell(t^\prime)$ conditional on all other random functions. Unlike \citet{qiao2019} which assumes a constant graphical model over the functional domain, we assume the graphical model varying over $t$, i.e. $G(t)=(V,E(t))$ with 
\bse
E(t) := \{(j, \ell): j\neq \ell, \; C_{j\ell}(t,t) \neq 0\}. 
\ese
Suppose we can represent each observed function sufficiently well by a truncated series of independent basis functions in the form
\be
y_{ij}(t) \approx \sum^K_{k=1} \phi_k(t) y^*_{ijk}=\by^*_{ij}\bdPhi(t),   \label{eq:basis_trans}
\ee
where $\bdPhi(t)=[\phi_1(t),\ldots,\phi_K(t)]'$ is the vector of basis functions, and $\by^*_{ij}=(y^*_{ij1},\ldots,y^*_{ijK})$ represents the row vector of the corresponding basis coefficients for the $(i,j)\th$ curve. The superscript $^*$ here (and hereafter) is used to denote the basis space parameters. Assume that this basis representation is \textit{lossless} or approximately lossless, i.e. $||y_{ij}(t)-y^*_{ij}\boldsymbol\Phi(t)||<\epsilon $ for some small value of $\epsilon$, for all $i$ and $j$, ensuring at least a vast majority of the total energy in all the observed functions is preserved by the basis representation; see \citet{Morris2011} for a more detailed discussion on lossless and near lossless basis representations. We will show later that this near lossless basis representation is sufficiently flexible to estimate functionally-evolving graphical models using our proposed model. The structure of the functional data will induce a particular within-function covariance structure across $t$. Here we will capture this by modeling basis coefficients as independent graphs. With suitable choice of basis, this can account for the salient features of within-function covariance including heteroscedasticity and various degrees of autocorrelation potentially varying across $t$, without requiring an unstructured covariance representation that is typically infeasible in this context. With the basis function representation, the conditional cross-covariance function is 
\be
C_{j\ell}(t,t^\prime) &\approx & Cov \left\{\sum^K_{k=1} \phi_k(t) Y^*_{jk},\sum^K_{k=1} \phi_k(t^\prime) Y^*_{\ell k} \; \Big| \; \sum^K_{k=1} \phi_k(s) Y^*_{hk},  \; h\neq j,\ell, \; \forall s \in \mathcal{T}\right\} \nonumber \\
&=& \sum_{k=1}^K \phi_k(t)\phi_k(t^\prime) \; Cov(Y^*_{jk},Y^*_{\ell k} \; | \; Y^*_{hk}, \; h\neq j,\ell ) \label{eq:ccc}
\ee
assuming independence of the basis coefficients across $k$, i.e. $Y^*_{jk} \perp Y^*_{h k^\prime}$ for all $k\neq k^\prime$. Thus we can induce conditional dependency between random functions by modeling the conditional dependency in the dual basis space. 

We propose a functional graphical model for inference of $G(t)$ which characterizes the functions using basis representation and models the dependency in the dual basis space. The method captures both within- and between-function correlations. Specifically our approach (i) utilizes basis function representations that model within-functional correlations and project the functional data into a basis space, (ii) constructs Bayesian Gaussian graphical models in the basis space using shrinkage priors that lead to induced adaptive shrinkage in data space, (iii) conducts Bayesian computation in the basis space, generating posterior samples that are then transformed back to the data space for inference of graphical models in  data space that change over the functional domain.

\underline{Basis transformation approach:}  In practice, $\{Y_{j}(t)\}_{j=1,\ldots,p}$ are observed only at a finite set of positions. Let $\by_{ij}$ be the row vector consisting of the $T$ observed values on a common grid, $t=1,\ldots,T$, within the interval $\mathcal{T}$, and $\by_i$ be the $p \times T$ matrix corresponding to the $p$ observed discrete functions from sample $i$, with the rows to be $\by_{ij}$. A discrete version of Equation (\ref{eq:basis_trans}) can be written as
$ 
\by_i = \by^*_i \bdPhi,
$ 
where $\bdPhi$ is the $K \times T$ matrix of basis functions evaluated at the $T$ observed positions. 
The matrix of basis coefficients in the dual basis space can be obtained by right-multiplying each side by the Moore-Penrose generalized inverse matrix of $\bdPhi$:
$ 
\by^*_i = \by_i \bdPhi'(\bdPhi \bdPhi')^{-1}.
$ 

This basis transformation approach can involve any generic basis functions $\phi$, such as functional principal components, wavelets, Fourier bases, or splines. In each of the cases, different algorithms can be used to calculate $\by^*$. For example, we can use the DWT algorithm in wavelet analysis, the fast Fourier transformation in Fourier analysis, and the singular value decomposition in PCA. Details of these strategies are discussed in \citet{Zhang2016fcar}. The choice of basis functions can be pre-determined by the characteristics of the application (e.g. wavelets, B-splines, Fourier basis) or determined empirically from the data (e.g. principal components). In particular, wavelets are suitable for irregular functions with spiky signals or discontinuities; Fourier bases are ideal for functions with stationary periodic features; and principal components work for sparse and smooth functional observations \citep{aston2010}. 

\underline{Basis-space graphical model:} We assume that for each basis function $k$, the $p$-dimensional basis coefficient vector (i.e. the $k\th$ column of $\by^*_i$) follows an independent multivariate Gaussian distribution, i.e.
\be
\by^*_{ik} & \sim & \mathcal{N} \left( \0, \{\Omega^*_k\}^{-1} \right),  \label{eq:wavelet_cov}
\ee
where $\Omega^*_k = [\omega^*_{k,jl}]_{p \times p}$ is a $p \times p$ precision matrix of the coefficient vector for the $k\th$ basis function. Note that the conditional covariance of the two basis coefficients, $Y^*_{jk}$ and $Y^*_{\ell k}$ can be derived by inverting their corresponding $2\times 2$ submatrix of $\Omega^*_k$, i.e.
\bse
Cov(Y^*_{jk},Y^*_{\ell k} \; | \; Y^*_{hk}, \; h\neq j,\ell) & = &  -\frac{\omega^*_{k,j\ell}}{\omega^*_{k,jj}\omega^*_{k,\ell\ell}-(\omega^*_{k,j\ell})^2 }. 
\ese
Given the conditional cross-covariance function (\ref{eq:ccc}) derived under the independence assumption of basis coefficients, the model specification results in an induced nonstationary Gaussian process in the data space with evolving conditional cross-covariances over the functional domain. In addition, the construction of graphical models in the basis space enables our functional graphical model to pool information across the functional domain, both locally and globally.  

We reparameterize the precision matrices as 
\bse
\Omega^*_k = [ \omega^*_{k,j\ell} ]_{p\times p} = D_{s^*_k}P^*_kD_{s^*_k}
\ese
where $D_{s^*_k}$ is a diagonal matrix with positive diagonal entries $s^*_{kj} = (\omega^*_{k,jj})^{1/2}$ and $P^*_k$ is positive definite matrix with diagonal entries of $1$ and off-diagonal entries $\rho^*_{k,ij}=\omega^*_{k,j\ell}/(\omega^*_{k,jj}\omega^*_{k,\ell\ell})^{1/2}$. This reparameterization follows Barnard et al. (2000) and Talluri et al. (2014), which separates the partial standard deviations and partial correlations. Now the conditional covariance of $Y^*_{jk}$ and $Y^*_{\ell k}$ becomes
\be
Cov(Y^*_{jk},Y^*_{\ell k} \; | \; Y^*_{hk}, \; h\neq j,\ell) 
& = & (s^*_{kj}s^*_{k\ell})^{-1} \left\{ - \frac{\rho^*_{k,j\ell}}{1-(\rho^*_{k,j\ell})^2}\right\}. \label{eq:cond_cov}
\ee

\begin{figure}
\centering
\centerline{\includegraphics[width=5in]{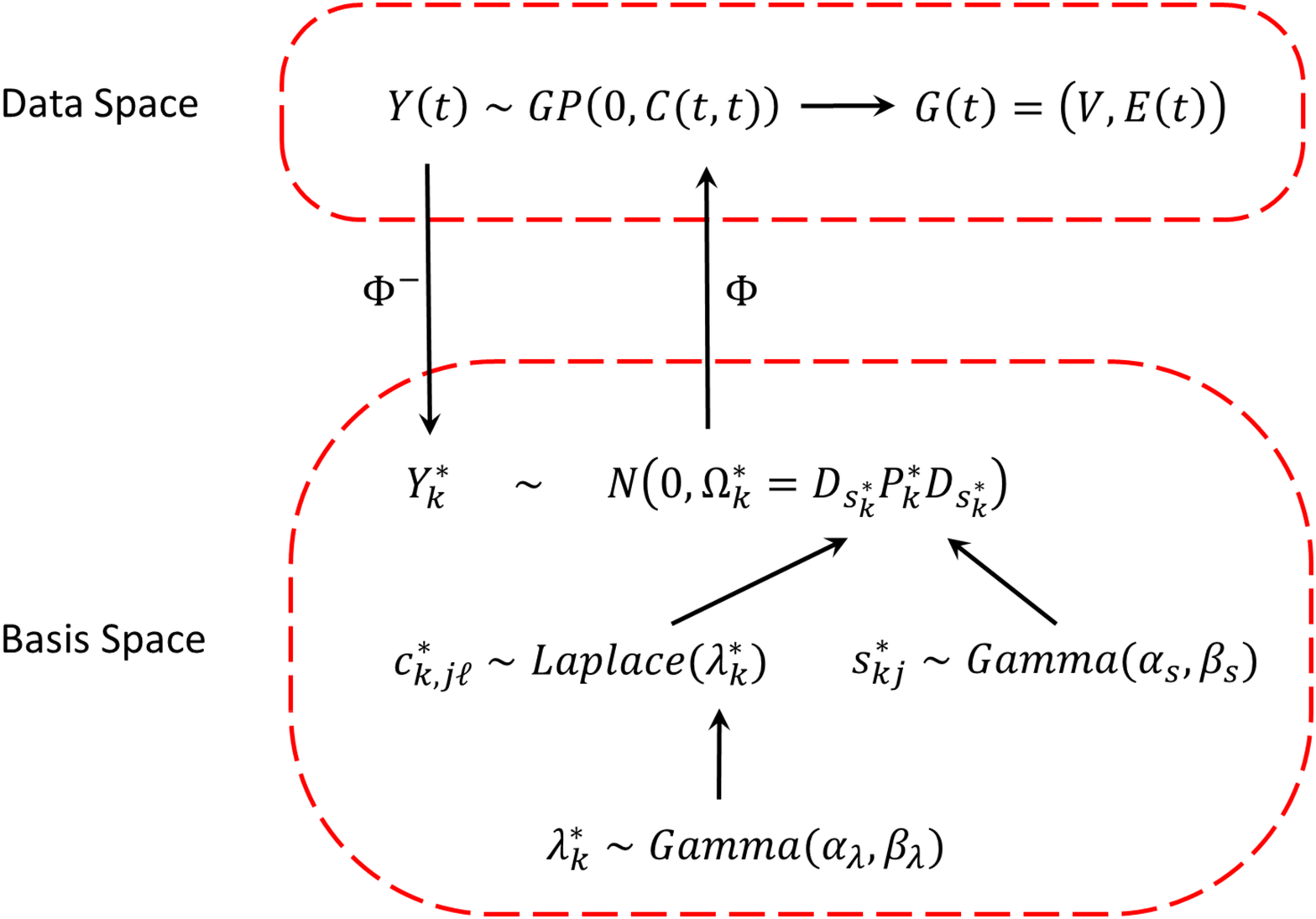}}
\vspace{.1in}
\caption{Schematic illustration of the Bayesian functional graphical model using basis function representation.}
\label{fig:full_model}
\end{figure}

We then use a variant of the Bayesian graphical lasso method to achieve shrinkage estimation of the basis-space precision matrix $\Omega^*_k$. In particular, we assign the following priors for $s^*_{kj}$ and transformed $\rho^*_{k,ij}$:
\bse
& s^*_{kj} & \sim \; Gamma(\alpha_s, \beta_s), \\
& c^*_{k,j\ell} = -\frac{\rho^*_{k,j\ell}}{1-(\rho^*_{k,j\ell})^2} & \sim \; Laplace(\lambda^*_k) \;  \mathcal{I}_{ P^*_k \in M^+}, \\
& \lambda^*_k & \sim \; Gamma(\alpha_\lambda, \beta_\lambda)
\ese
where $(\alpha_s, \beta_s)$ are specified for a vague gamma prior on $s^*_{kj}$, the indicator $\mathcal{I}_{ P^*_k \in M^+}$ constraints $c^*_{k,j\ell}$ so that $P^*_k$ is positive definite, and $\lambda^*_k$ is the $k$-specific regularization parameter with a vague gamma prior for joint estimation. Compared to the Bayesian graphical lasso that applies an $L_1$ shrinkage prior on the partial correlations, our Bayesian model shrinks the transformed parameter $c^*_{k,j\ell}$, which leads to shrinkage on the conditional covariance of basis coefficients and consequently the conditional cross-covariance matrices in the data space as we will show later. We choose the Laplace prior as in typical Bayesian graphical lasso because it induces a novel normal-gamma type of shrinkage prior for the conditional cross-covariance matrices in the data space, as will be shown in Sections \ref{subsec:prior} and \ref{subsec:prior_prop}, that facilitates a thorough examination of its adaptive shrinkage property and robust tail behavior. However, other sparsity/shrinkage priors could also be potentially considered, for example, the normal-gamma prior \citep{griffin2010}, the horseshoe prior \citep{carvalho2010}, the generalized double Pareto prior \citep{armagan2013}, and the Dirichlet-Laplace prior \citep{bhattacharya2015}, which we expect to lead to induced priors in data space with similar properties.  An illustration of the Bayesian functional graphical model is shown in Figure \ref{fig:full_model}.

\subsection{Induced Nonstationary Gaussian Process in Data Space} \label{subsec:likelihood}

We first present lemmas that formally relate the basis-space graphical model to an induced nonstationary Gaussian process in data space. Suppose we have $n$ subjects, and each subject $i$ has $p$ functional curves $\{Y_{ij}(t)\}_{1,\ldots,p}$ corresponding to $p$ random variables, where each function $Y_{ij}(t)$ is observed on a common set of $T$ discrete positions. Consider the following assumptions: \\

{\noindent(A1)} Suppose we define a basis function matrix $\boldsymbol\Phi=[\phi_k(t)]_{K\times T}$ of dimensionality $K \times T$ $(K \leq T)$  that is of full row rank, i.e., rank$(\boldsymbol\Phi)=K$, which transforms the observed functions into the basis space by $\bY^*_i=\bY_i\bPhi^\prime(\bPhi\bPhi^\prime)^{-1}$. \\

\textsc{LEMMA 1}: {\it Under assumption (A1), if $\bY^*_{ik}=(Y^*_{i1k},\ldots,Y^*_{ipk})'$ is independently Gaussian-distributed with the precision matrix $D^*_{s_k}P^*_kD^*_{s_k}$ for $k=1,\ldots, K$, the vector of random functions $\bY_{i}(t) =(Y_{i1}(t),\ldots,Y_{ip}(t))'$ follows an induced nonstationary Gaussian process with the conditional cross-covariance matrix function:
\bse
C(t,t) = [ C_{jl}(t,t) ]_{p \times p}, \;\; C_{jl}(t,t) = \sum^K_{k=1} \phi_k^2(t)  (s^*_{kj}s^*_{k\ell})^{-1} c^*_{k,j\ell} \; , 
\ese
where $c^*_{k,ij} = -\rho^*_{k,ij}/\{1-(\rho^*_{k,ij})^2\}$.
} \\

\textsc{LEMMA 2}: {\it Under assumption (A1), if $\bY^*_{ik}=(Y^*_{i1k},\ldots,Y^*_{ipk})'$ is independently Gaussian-distributed with the precision matrix $D^*_{s_k}P^*_kD^*_{s_k}$ for $k=1,\ldots, K$, the random functions at locations $t$ and $'t^\prime$, $\bY_{i}(t)$ and $\bY_{i}(t')$, have the conditional cross-covariance matrix as
\bse
C(t,t^\prime) = [ C_{jl}(t,t^\prime) ]_{p \times p}, \;\; C_{jl}(t,t^\prime) = \sum^K_{k=1} \phi_k(t)\phi_k(t^\prime)  (s^*_{kj}s^*_{k\ell})^{-1} c^*_{k,j\ell} \; .
\ese
}

{\bf Remarks:} Lemmas 1\&2 are direct results of (\ref{eq:ccc}) and (\ref{eq:cond_cov}). Lemma 1 provides the conditional dependency among the $p$ variables at each $t$ in the induced Gaussian process. Lemma 2 provides information on whether and how the variables are correlated between any two different locations. More importantly, compared to existing time-varying Gaussian graphical models which typically treat the observed data at each location $t$ as independent, our functional graphical model accounts for within-function correlations and pools information from both local and remote regions in graphical modeling.

\subsection{The Induced Shrinkage Prior in Data Space} \label{subsec:prior}

We now examine the induced shrinkage priors for $C_{j\ell}(t,t)$ using our functional graphical model and their properties. Our Bayesian model assumes independent Laplace priors on $c^*_{k,j\ell}$, and thus by Lemma 1 the conditional cross-covariance $C_{j\ell}(t,t)$ in the data space are weighted sums of independent Laplace random variables. We will show that this construction leads to a new class of  normal-scale-mixture shrinkage priors, the  {\it normal-hypoexponential distribution}, for $C_{j\ell}(t,t)$.

\textbf{Normal-hypoexponential distribution:} \citet{smaili2013} showed that the sum of independent exponential random variables has a valid continuous distribution over the positive real line and defines it to be a hypoexponential distribution.

\textit{The hypoexponential distribution:} Let $X_1, \ldots, X_K$ be independent exponential random variables with different respective parameters $\lambda_k$, $k=1,\ldots,K$, written as $X_k \sim \mathrm{Exp}(\lambda_k)$, then the sum of the random variables, $X=\sum^K_{k=1}X_k$ has a {\it hypo-exponential} distribution with parameters $\lambda_1,\ldots,\lambda_K$, written as
\bse
X \sim \mathrm{Hypo}(\lambda_1,\ldots,\lambda_K).
\ese
The hypoexponential distribution has the following probability density function (pdf):
\be
f(x) = \sum^K_{k=1} \left\{ \prod_{h \neq k}(1-\lambda_k/\lambda_h)^{-1} \right\} \lambda_k e^{-\lambda_k x},  \label{eq:hypo_pdf1}
\ee
with mean $E(X) = \sum^K_{k=1} \frac{1}{\lambda_k}$, and variance $\mathrm{Var}(x) = \sum^K_{k=1} \frac{1}{\lambda_k^2}$. We see that the pdf of the hypoexponential distribution is actually a linear combination of the pdfs of $K$ independent exponential distributions. We can rewrite the pdf (\ref{eq:hypo_pdf1}) as
\be
f(x) = \sum^K_{k=1} P_k f_k(x|\lambda_k),  \label{eq:hypo_pdf2}
\ee
where $f_k(\cdot)$ is the pdf of an exponential distribution, and $P_k = \prod_{h \neq k}(1-\lambda_k/\lambda_h)^{-1}$ is the coefficients for the $k\th$ exponential component. 


Figure S1 in the supplementary materials displays the density curves of two hypoexponential distributions from our real data analysis in Section \ref{sec:app}. The hypoexponential is a gamma-type distribution. When one rate parameter is much smaller than all others, $X$ is approximately to be exponentially distributed. In the other extreme case when $\lambda_1 = \cdots=\lambda_K=\lambda$, $X$ becomes a gamma random variable.

The following lemma shows that the sum of independently Laplace-distributed random variables follows a normal-hypoexponential (N-Hypo) scale mixture distribution. 

\textsc{LEMMA 3}: {\it Let $X_1, \ldots, X_K$ be independent Laplace random variables with different respective parameters $\lambda_k$, $k=1,\ldots,K$, written as $X_k \sim \mathrm{Laplace}(\lambda_k)$, then the sum of the random variables, $X=\sum^K_{k=1}X_k$ has a {\it normal scale mixture distribution with the mixing distribution to be hypoexponential} with parameters $\lambda_k/2$, $k=1,\ldots,K$. Mathematically,
\bse
X & \sim & \mathcal{N}(0,\tau), \\
\tau &\sim & \mathrm{Hypo} \left( \frac{\lambda_1}{2},\ldots,\frac{\lambda_K}{2} \right).
\ese
}
\textit{Proof.} See Appendix.

\textbf{Induced shrinkage priors for $C_{j\ell}(t,t)$:} By definition of hypoexponential distribution and Lemma 3, we obtain that the functional graphical model induces a N-Hypo prior in the data space for the conditional cross-covariance $C_{jl}(t,t) = \sum^K_{k=1} \phi_k^2(t)  (s^*_{kj}s^*_{k\ell})^{-1} c^*_{k,j\ell}$:
\bse
C_{jl}(t,t)  \sim  \mathcal{N}(0,\tau_{jl}),  \quad
\tau_{jl} \sim  \mathrm{Hypo} \left( \frac{\lambda_{t1}}{2},\ldots,\frac{\lambda_{tK}}{2} \right) \; , 
\ese
where $\lambda_{tk} = \lambda^*_k s^*_{kj}s^*_{k\ell}/\phi^2_k(t)$ for $k=1,\ldots,K$. The hypoexponential distribution is a gamma-type distribution, and hence the induced N-Hypo prior shrinks $C_{j\ell}(t,t)$ toward zeros. In the extreme case when $\phi^2_{k}(t)=1$ for some $k$ and $0$ for all other bases, we have $\tau_{jl} \sim \mathrm{Exp}(\lambda^*s^{*2}_k)$, leading to a normal-exponential prior or a lasso shrinkage of $C_{j\ell}(t,t)$.

\subsection{Properties of the N-Hypo prior} \label{subsec:prior_prop}

We use a simple case to examine the shrinkage properties of the N-Hypo prior. Suppose we observe $n$ samples $y=(y_1, \ldots, y_n)$ of one single random variable $Y|\mu \sim \mathcal{N}(\mu,\sigma^2)$, and that our interest is to estimate $\mu$. We propose to estimate $\mu$ using the posterior mean with the N-Hypo mixture prior as follows:
\be
\begin{array}{rcl}
\mu | \sigma^2, \tau^2 & \sim & \mathcal{N}(0,\sigma^2\tau^2), \\
\tau^2 | \lambda_1, \ldots,\lambda_K & \sim & \mathrm{Hypo}(\frac{\lambda_1}{2}, \ldots, \frac{\lambda_K}{2}). 	
\end{array}	\label{eq:norm-hypo}
\ee

\textbf{Tail robustness:} First we look at the tail behavior of the normal mixture prior (\ref{eq:norm-hypo}), i.e. the shrinkage for large signals. Based on the work of \citet{pericchi1992}, we can represent the posterior mean of $\mu$ as
\bse
E(\mu|\bar{y}) = \bar{y} - \frac{\sigma^2}{n} S(\bar{y}), \label{eq:shrink}
\ese
where $\bar{y}$ is the mean of the $n$ observations, and 
\bse
S(\bar{y}) & = & -\frac{d}{d\bar{y}}\log m(\bar{y}), 
\ese
where $m(\bar{y}) = \int \mathcal{N}(\bar{y}|\mu,\sigma^2/n)\pi(\mu)d\mu$ is the predictive distribution of $\bar{y}$. Therefore, the shrinkage size of a prior on $\mu$ is given by $-\frac{d}{d\bar{y}}\log m(\bar{y})$, and $\lim_{\bar{y} \rightarrow \infty} S(\bar{y})$ gives its shrinkage behavior for large signals. Without loss of generalization, we set $\sigma^2=1$ for the rest of the section. 

\textsc{Theorem 1:} {\it Suppose $\bar{y} \sim \mathcal{N}(\mu,1/n)$. Let $m_k(\bar{y})$ denote the predictive density under the normal-exponential prior with a scale parameter $\lambda_k$, i.e. where $\mu \sim \mathcal{N}(0,\tau^2)$ and $\tau^2 \sim \mathrm{Exp}(\lambda_k/2)$, and $S_k(\bar{y}) = -\log m_k(\bar{y})/d\bar{y}$. Then 
\bse
\lim_{\bar{y} \rightarrow \infty} S_k(\bar{y}) = \sqrt{\lambda_k} \; .
\ese
\\}
\textit{Proof.} See Appendix. 

\textsc{Theorem 2:} {\it Suppose $\bar{y} \sim \mathcal{N}(\mu,1/n)$. Let $m(\bar{y})$ denote the predictive density under the N-Hypo prior, i.e. where $\mu \sim \mathcal{N}(0,\tau^2)$ and $\tau^2 \sim \mathrm{Hypo}(\lambda_1/2,\ldots,\lambda_K/2)$, and $S(\bar{y})  =  -\log m(\bar{y})/d\bar{y}$. Then 
\bse
\lim_{\bar{y} \rightarrow \infty} S(\bar{y}) = \sqrt{\lambda_\kappa} \; ,
\ese
where $\kappa = \arg\!\min_k \{\lambda_k: k=1,\ldots,K \}$.  
\\}
\textit{Proof.} See Appendix. 

Theorems 1\&2 show that the tail behavior of a N-Hypo prior is determined by and equivalent to that of the exponential component with the smallest shrinkage (smallest $\lambda_k$). In our functional graphical model settings, by Lemma 1 $C_{j\ell}(t,t)$ has an induced N-Hypo prior with parameters $\{\lambda_{tk}/2, k=1,\ldots,K \}$, $\lambda_{tk} = \lambda^*_ks^*_{kj}s^*_{k\ell}/\phi^2_{k}(t)$. Note that $\lambda_{tk}$ becomes infinite when the basis function $\phi^2_{k}(t)$ is zero at $t$. Therefore, the tail robustness of the induced N-Hypo prior at position $t$ in the data space is only determined by those basis functions with significantly nonzero values at $t$ (i.e. $\phi^2_{k}(t)>>0$). More specifically, if two variables have high conditional covariance for at least one basis function $k$ with $\phi^2_{k}(t)>>0$ for some $t$, their conditional cross-covariance will be high at $t$ in the data space, indicating a strong conditional dependency between them. Also note that $(s^*_{kj})^{-2}$ is the conditional variance of $Y^*_{jk}$ given $\{Y^*_{\ell k}: \ell \neq j\} $, which is smaller than its marginal variance. This means that basis functions with small values of coefficients $y^*_{ik}$ (and thus small variances and large $s^*_{kj}$) will have a large value of $\lambda_{tk}$ and do not impact the tail behavior of the normal-hypo prior. The tail robustness of the shrinkage prior is predominantly determined by those basis functions that have a great contribution to the total energy of the observed functions, which justifies our near lossless basis representation of the observed data.

\textbf{Shrinkage around zero:} We now look at the shrinkage property of the N-Hypo around zero. The shrinkage strength for weak signals is determined by the mass of the scale mixing distribution close to zero, i.e. $P(\tau^2<\epsilon)$ for a small value $\epsilon$. For the scale mixing distribution Hypo($\lambda_1, \ldots, \lambda_K$), we have
\bse
P(\tau^2<\epsilon) & = & \int^\epsilon_0 \ \sum_{k=1}^K P_k \frac{\lambda_k}{2} e^{-\lambda_k\tau^2/2} \ d\tau^2 \\
		&=& 1 - \sum_{k=1}^K P_k e^{ -\frac{\lambda_k}{2}\epsilon }.
\ese
A sufficient condition for strong shrinkage around zero, i.e. $P(\tau^2<\epsilon) \rightarrow 1$, is that $\lambda_k \rightarrow \infty$ for all $k$. In our model where the N-Hypo parameters for $C_{j\ell}(t,t)$ are $\lambda_{tk}/2$ with $\lambda_{tk} = \lambda^*_ks^*_{kj}s^*_{k\ell}/\phi^2_{k}(t)$, the sufficient condition is equivalent to $\lambda^*_k \rightarrow \infty$ for all bases $k$ of bounded $s^*_{kj}$ and $\phi^2_{k}(t) >> 0$. This implies that the shrinkage strength around zero for $C_{j\ell}(t,t)$ is jointly determined by basis functions that explain a considerable proportion of data variances and significantly nonzero at $t$.

\textbf{Induced shrinkage properties for the functional graphical models:}
The degree of shrinkage at a certain $t$ depends on the basis functions supported at $t$, the shrinkage parameters for those basis functions, and the conditional variance of the basis coefficients. Thus, the set of locations $t$ having high magnitude for the same basis function $k$ tend to have similar degrees of sparsity. For wavelet basis, the support of high frequency wavelets are concentrated locally, while that of low frequency wavelets are more global. For functional principal components, the range of support for the eigenfunctions could be local, global, or distant. Thus, this prior is able to borrow strength across $t$ in the shrinkage both locally and globally, according to the chosen basis functions. This also ensures that the resulting graphical models inferred in the data space evolve smoothly over the functional domain.

\section{Posterior Inference} \label{sec:postinf}


We derive the full conditional distributions of the parameters $(\mathbf{s}^*,\mathbf{c}^*, \boldsymbol\lambda^*)$ in the dual basis space and use a block Gibbs sampling algorithm to generate posterior samples of them.

\begin{itemize}
\item Sampling of $c^*_{k,j\ell}$: \\
Instead of sampling $c^*_{k,j\ell}$ directly, we sample the partial correlation $\rho^*_{k,j\ell}= \frac{1-\sqrt{1+4c^{*2}_{k,j\ell}}}{2c^*_{k,j\ell}}$ and then transform to $c^*_{k,j\ell}$. Let $P^*_k = R^TR$ be the Cholesky decomposition of $P^*_k$ where the matrix $R$ is upper triangular. Without loss of generality, suppose that $j=p-1$ and $\ell = p$. The full conditional distribution of $\rho^*_{k,j\ell}$ is 
\bse
p(\rho^*_{k,j\ell}|\cdot) & \propto & |P^*_k|^{n/2} \exp \left\{-\frac{1}{2}tr\left(\by^{*T}_kD_{s^*_k}P^*_kD_{s^*_k} \by^*_k \right) -\frac{\lambda_k \; |\rho^*_{k,j\ell}|}{1-\rho^{*2}_{k,j\ell}} \right\} (1+\rho^{*2}_{k,j\ell})(1-\rho^{*2}_{k,j\ell})^{-2} \\
& \propto & \left\{ 1-\left( \frac{\rho^*_{k,j\ell}-a}{b} \right)^2 \right\}^{n/2} \exp \left\{-\Lambda_{k,j\ell} \; \rho^*_{k,j\ell} -\frac{\lambda_k \; |\rho^*_{k,j\ell}|}{1-\rho^{*2}_{k,j\ell}} \right\} \\
&& \times \; (1+\rho^{*2}_{k,j\ell})(1-\rho^{*2}_{k,j\ell})^{-2} \cdot \mathcal{I}(|\rho^*_{k,j\ell}-a|<b)
\ese
where $\Lambda_k=D_{s^*_k} \by^*_k\by^{*T}_kD_{s^*_k}$, and $a=\sum_{r=1}^{p-2}R_{r,p-1}R_{r,p}$ and $b=\sqrt{R_{p-1,p-1}(R^2_{p-1,p}+R^2_{p,p})}$ do not depend on $\rho^*_{k,j\ell}$. The positive definite constraint on $P^*_k$ is ensured by the indicator function at the end. The full conditional is not in closed form and cannot be sampled directly. Since the density has support only over $(a-b,a+b)$, we use an independent Metropolist-Hastings (MH) algorithm to sample $\rho^*_{k,j\ell}$. We choose $100$ equally spaced grids spanning the interval, calculate the densities at the grids, and construct a piecewise uniform distribution with jumps at the grids as our independent MH proposal density. This piecewise uniform proposal distribution well approximates the target full conditional, and therefore MCMC chains using the independent MH algorithm have a high acceptance rate and rapid convergence compared to a random-walk MH algorithm. The posterior samples of $c^*_{k,j\ell}$ are then obtained as $-\rho^*_{k,j\ell}/(1-\rho^{*2}_{k,j\ell})$.

\item Sampling of $s^*_{kj}$: \\
The full conditional distribution of $s^*_{kj}$ is
\bse
p(s^*_{kj}|\cdot) & \propto & \left( s^*_{kj} \right)^{n+\alpha_s-1}\exp \left\{ -\frac{1}{2} \Gamma_{k,jj}s^{*2}_{kj}- \left( \sum_{\ell:\ell \neq j} \Gamma_{k,j\ell}s^*_{k\ell}\rho^*_{k,j\ell} + \beta_s \right)s^*_{kj} \right\},
\ese
where $\Gamma_k=\by^*_k\by^{*T}$. The distribution does not have a closed from. We use a MH algorithm to draw $s^*_{kj}$.

\item Sampling of $\lambda^*_k$: \\
The full conditional distribution of $\lambda^*_k$ is
\bse
p(\lambda^*_{k}|\cdot) & \sim & Gamma \left(\alpha_\lambda+\frac{p(p-1)}{2}, \beta_\lambda+\sum_{j < \ell}  c^*_{k,j\ell} \right),
\ese
and posterior samples of $\lambda^*_k$ can be directly drawn from a gamma distribution.

\end{itemize}

These posterior samples of $(\mathbf{s}^*,\mathbf{c}^*)$ are then transformed back into the data space, yielding posterior samples of the conditional cross-covariance $C_{j\ell}(t,t)$, based on which 95\% credible intervals are constructed. Then the estimated graphical models are obtained as 
\bse
\widehat{E}(t) = \{ \; (j,\ell): j \neq \ell, 0 \notin (LB_{C_{j\ell}(t,t)}, UB_{C_{j\ell}(t,t)}) \; \} \; ,
\ese
where $LB_{C_{j\ell}(t,t)}$ and $UB_{C_{j\ell}(t,t)}$ are lower and upper bounds of the credible interval of $C_{j\ell}(t,t)$. Similar to single graphical modeling, our method is quadratic in $p$ in computational time. But note that the Bayesian computation in the basis space is linear in $K$ and can be parallelized due to the independence of the basis coefficients across $k$. The computation parallelizability joint with near-lossless basis representation (with $K << T$) allows the method to scale up to high-dimensional multi-variate functional data of large $p$ or $T$.

\section{Simulation Studies} \label{sec:simu}

\subsection{Simulation setting}

\begin{figure}
\centering
\centerline{\includegraphics[height=3in]{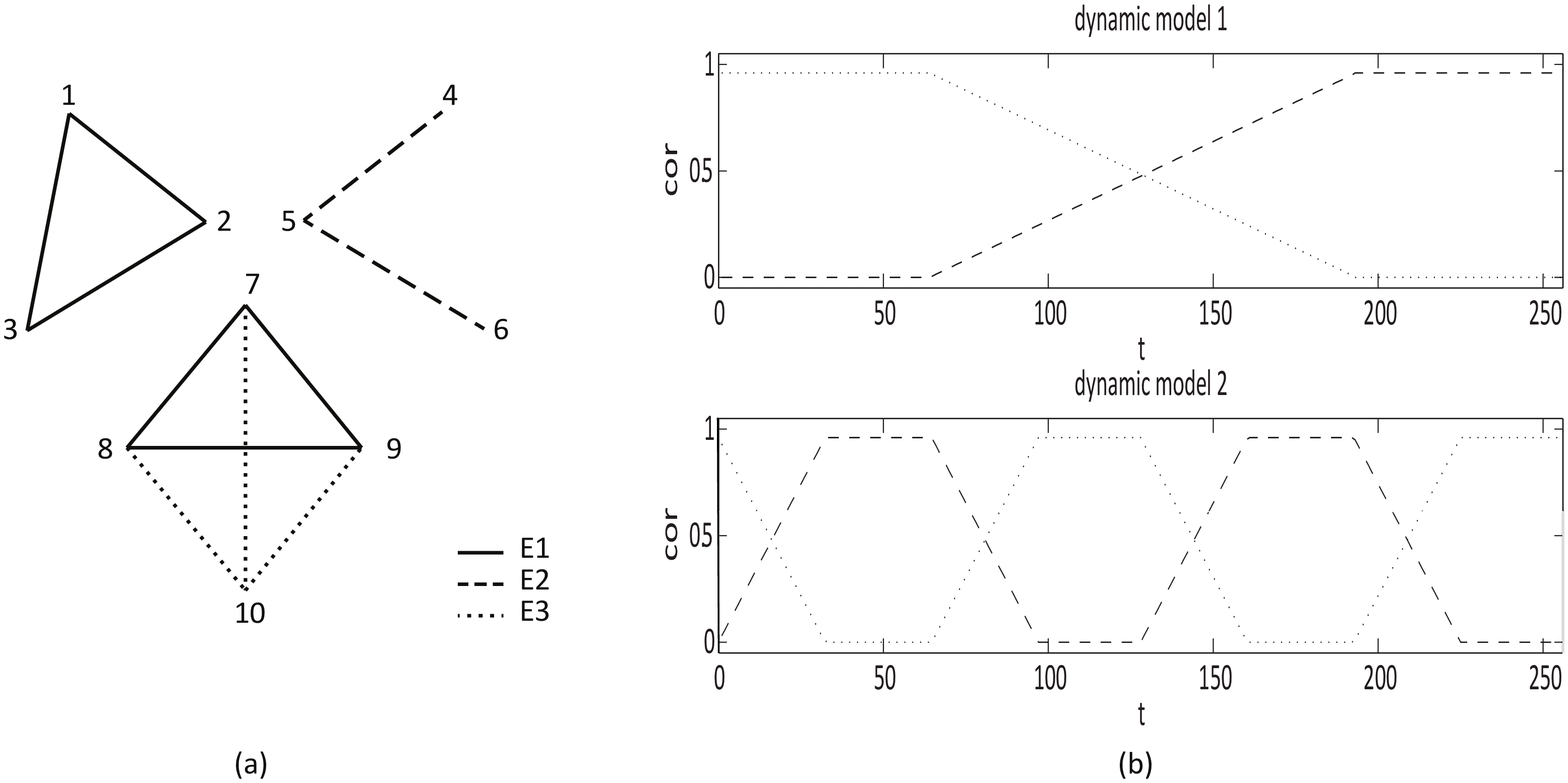}}
\caption{The Gaussian graphical model in the simulation study. Edge set E1(solid lines) corresponds to connections with constant correlations across $t$; Edge sets E2 (dashed) and E3 (dotted) correspond to connections with correlations that vary across time as shown in (b).}
\label{fig:simu_graph}
\end{figure}

In this section, we present results from a simulation study designed to examine the performance of the Bayesian functional graphical model for estimating functionally-evolving network structures. We considered two autocorrelation scenarios from which multivariate functional data were generated:
\bse
\text{AR(1) model:} & \bY(t)  &  =  A \bY(t-1) + \bepsilon(t), \\
\text{Change-point model:} & \bY(t)  & =  A_1 \bX_1 I(t <= t_0) + A_2 \bX_2 I(t > t_0) + \bepsilon(t),
\ese
where $\bY(t)$ is a $p-$dimensional random vector observed at functional position $t$, $A$, $A_1$, and $A_2$ are regression coefficient matrices for the AR(1) and Change-point models respectively that bring correlations across $t$, and $\bepsilon(t)$ is a vector of residuals with a Gaussian distribution $\mathcal{N}_{p}(\0, \Sigma_t)$, which evolves over the functional domain. The AR(1) model represents the cases when within-functional autocorrelations are between nearby functional locations, and the change-point model represents multivariate functional data types with long-range within-functional autocorrelations. In our simulation study, we specify $p=10$, and $A$, $A_1$, $A_2$ all to be diagonal matrices. In addition, we assume that the residual vector $\bepsilon(t)$ follows a sparse Gaussian graphical model as shown in Figure \ref{fig:simu_graph}(a). To allow for the sparse graphical model evolving over the functional domain, we let the correlations corresponding to two edge subsets, E2 and E3, in the figure changing across $t$. For each of autocorreltion scenarios, we considered two functionally-evolving graphical models of the residuals as shown in Figure \ref{fig:simu_graph}(b), in each of which the cross-correlations of edge subsets E2 and E3 vary continuously between 0 and 1, resulting in sparse graphical structures that change over the functional domain. The graphical models at various positions for dynamic model 1 is more clearly illustrated by plots that are available in Figure S2 in the supplementary materials.

\subsection{Simulation result}

We generated 100 datasets for each of the four settings with sample size $n=50$. For each simulated dataset, we applied our Bayesian functional graphical models using Daubechies 2 wavelet bases, the Bayesian independent graphical models which uses Bayesian graphical lasso for estimation at each location separately, and the fused graphical lasso that encourages local smoothness by penalizing neighoring differences. For the Bayesian functional and indepedent graphical models, we used 95\% credible intervals for posterior inference of significant edges based on $1000$ MCMC samples with a thinning of 5 and $1000$ burn-in iterations. The tuning parameters for the fused graphical lasso method were chosen by the Bayesian information criterion (BIC). The performance of the models was evaluated by (i) the integrated mean true positive rate (IMTPR) summarizing the sensitivity of the graphical models in detecting the true edges, and (ii) the integrated mean false positive rate (IMFPR) summarizing the specificity in inference of the graphical structures averaged across $t$.

Table \ref{table:sim1} presented the means and standard deviations of IMTPRs and IMFPRs over the functional domain over all replications. We see that the Bayesian independent graphical model has lowest powers (IMTPRs) in detecting the true edges of all the methods, since it fails to borrow information across functional locations in graph estimation. The fused graphical lasso method has similar powers to our method for the scenarios with AR(1) autocorrelations but obviously lower powers for the change-point models. This is because the fused graphical lasso only uses local information by assuming similar graphs at nearby locations but does not account for long-range autocorrelations. The high IMFPRs also indicate overfitting of models selected by the BIC criteria. The Bayesian functional graphical model has the best performance with high IMTPRs and low IMFPRs for all scenarios. This suggests that our method is able to boost powers while controlling type 1 error rates by accounting for both short- and long-range within-functional correlations and pooling both local and global information across $t$.


\begin{table}
\caption{Inference of functionally-varying network structures by the wavelet-based Bayesian functional graphical models, Bayesian independent graphical models, and fused graphical lasso. Graph estimation performance is evaluated in terms of integrated mean true positive rates (IMTPRs) and integrated mean false positive rates (IMFPRs). The means with the standard deviations in the parenthesis across 100 replications are presented.}
\label{table:sim1} 
\fbox{%
\begin{tabular}{|cl|c|c|c|c|}
\hline
\multicolumn{2}{|c|}{} & \multicolumn{2}{c|}{AR(1) Model} & \multicolumn{2}{c|}{Change-Point Model}\\
\cline{3-6}
& & {Dynamic 1} & {Dynamic 2} & {Dynamic 1} & {Dynamic 2}\\
\hline
\parbox[t]{4mm}{\multirow{3}{*}{\rotatebox[origin=c]{90}{IMTPR}}}
& BayesFunc & 0.997 (0.003) & 0.998 (0.002) & 0.981 (0.017) & 0.901 (0.015) \\
& BayesInd & 0.839 (0.008) & 0.833 (0.008) & 0.732 (0.020) & 0.669 (0.011) \\
& FusGLasso & 0.978 (0.005) & 0.970 (0.006) & 0.926 (0.025) & 0.872 (0.022)\\
\hline
\parbox[t]{4mm}{\multirow{3}{*}{\rotatebox[origin=c]{90}{IMFPR}}}
& BayesFunc & 0.008 (0.004) & 0.010 (0.004) & 0.014 (0.008) & 0.013 (0.009) \\
& BayesInd & 0.019 (0.002) & 0.019 (0.002) & 0.012 (0.003) & 0.013 (0.003) \\
& FusGLasso & 0.062 (0.015) & 0.084 (0.023) & 0.115 (0.026) & 0.113 (0.025) \\
\hline
\end{tabular} }
\end{table}

\begin{figure}
\centering
\centerline{\includegraphics[height=5in,width=5in]{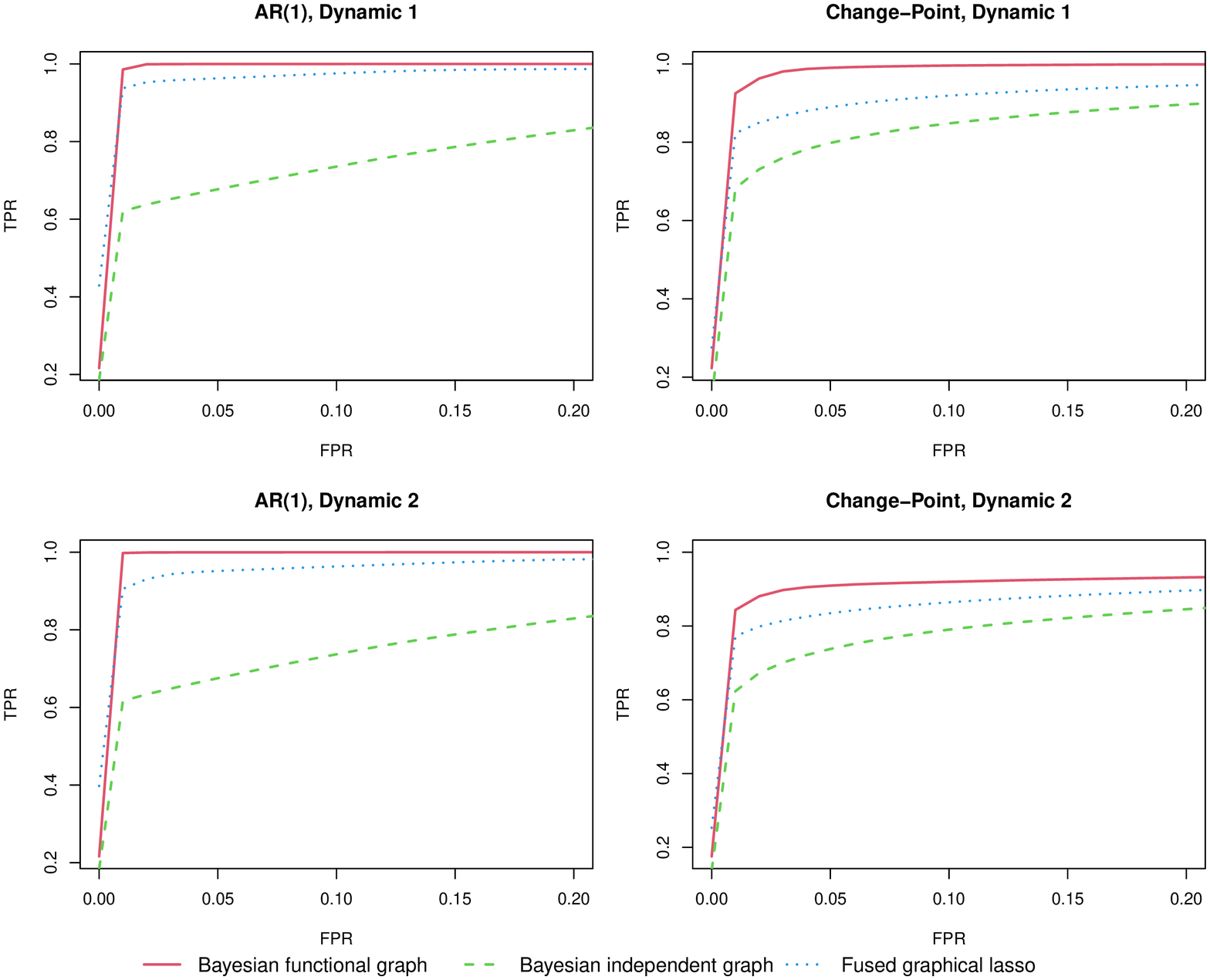}}
\caption{Average ROC curves of the Bayesian functional graphical model, Bayesian independent graphical model, and fused graphical lasso methods, over 100 replications. The ROC curves for the Bayesian methods were obtained by thresholding the posterior mean estimates of the conditional cross-covariance and precision matrices respectively; those for the fused graphical lasso were obtained by thresholding the estimated precision matrices selected by the BICs.}
\label{fig:simu_roc}
\end{figure}

To illustrate the performance of all methods more comprehensively, we present the receiver operation characteristic (ROC) curves averaged across 100 replications in Figure \ref{fig:simu_roc}. The ROC curves for the Bayesian functional graphical model and independent graphical model were obtained by thresholding the posterior mean estimates of the conditional cross-covariance and precision matrices respectively; those for the fused graphical lasso were obtained by thresholding the estimated precision matrices selected by the BICs. We see that the Bayesian independent graphical model has the worst and consistent performance for all four scenarios due to its failure to account for within-functional autocorrelations. The Bayesian functional graphical model and fused graphical lasso have similar performance for the two scenarios with AR(1) autocorrelations, but the latter has larger decreases in the area under the curve (AUC) values for the scenarios with change-point models. This is consistent with our observations in Table \ref{table:sim1} that the fused graphical lasso can only borrow local information while our method can account for both short- and long-range within-functional correlations and pool information from both local and global regions.

As indicated by Lemma 2, in addition to being able to identify the connections among variables at a certain position $t$, our Bayesian functional graphical model is also able to estimate the correlations between variables at two different locations, $t$ and $t'$, which can shed information on the pattern of within-functional autocorrelations. Figure S3 in the supplementary materials displays the estimated lagged conditional cross-covariance between Nodes 1\&2, $C_{12}(t,t^\prime)$, for $t=1$ and $t^\prime=2,\ldots,T$, averaged across 100 replications. The plots show a exponentially decay trend of the conditional cross-covariances for the two AR(1) settings as time lag increases and a piece-wise constant trend for the two change-point settings, which are consistent with the autocorrelation patterns of the simulated multivariate functional data. These results indicate that our Bayesian functional graphical methods are able to detect and provide information on the pattern of within-functional autocorrelations.

\section{Data Application} \label{sec:app}

\subsection{Data Description}
We applied our method to a task fMRI dataset collected from a recent study on neurofeedback (NF) intervention of adolescent depression \citep{quevedo2019}. Adolescence is a period of increased risk for developing depression and early onset is associated with a poorer prognosis, higher symptom severity along with higher suicidality rates. Heightened self-focus with rigid negative self-representations are found to converge with mental illness among depressed adolescents. Dr. Quevedo's group recently developed an intervention using NF training to enhance positive self-processing, which allows voluntary modulation of brain activity ``in vivo" during fMRI experiment. The NF-task fMRI experiment included four blocks of NF training, during which adolescents were asked to recall happy autobiographical memories to increase real-time monitored activities of amygdala area displayed via a colored bar shifting up or down while seeing their smiling face. Each NF-training block was followed by a control block during which participants counted backward from 100 during the control condition while seeing an unfamiliar face. The details of the NF-task designs are provided in \citet{quevedo2019}. Previous analyses show significant symptom reduction among depressed youths after NF training, and changes of depressed adolescents’ functional connectivity (FC) between relevant brain regions during NF-task fMRI is of key interest to indicate possible mechanisms for its effectiveness.

The fMRI dataset includes a total of $34$ depressed adolescents and $19$ healthy controls, each subject has $235$ volumes over an fMRI scan that lasts for $354$ seconds. We focus on $17$ regions of interest (ROIs) (listed in Table S1 in supplementary materials) that were found to be significantly activated during NF-task blocks, from which mean time series signals were extracted. Thus the final data for analysis have a dimension of $n=34$ or $19$, $p=17$, and $T=235$. All time series were centered at zero and data of each ROI were normalized with standard deviation of 1. We then applied our Bayesian functional graphical model to the normalized fMRI data of depressed adolescents and healthy controls separately, with the aim to detect the dynamic neurological activity in response to the stimuli of smiling self-faces versus other-faces during the NF training. In comparison, we also applied the Bayesian independent graphical model and the fused graphical lasso method. 

\subsection{Analysis Results}
We used the Daubechies 2 wavelet basis functions, periodic-padding boundary, and decomposed to 6 wavelet levels for basis transformation in our Bayesian functional graphical model. The Daubechies wavelets can capture the spiky signals that are often present in fMRI data. We specified vague Gamma$(0.1,0.1)$ hyperpriors for both $s^*_{kj}$ and $\lambda^*_k$. For the Bayesian functional and independent graphical models, we collected 2000 posterior samples with a thinning of 5 from the MCMC chain after a 5000 burn-in iterations. Figure S4 shows the traceplots of conditional cross-covariances of two identified edges at three time points from the Bayesian functional graphical model. The graphical models at each time point $t$ were then obtained using 95\% credible intervals of the conditional cross-covariances and the precision matrices respectively. The time-varying graphical model for the fused graphical lasso was selected by BIC values. 

\begin{figure}
\centering
\centerline{\includegraphics[height=5in,width=7.5in]{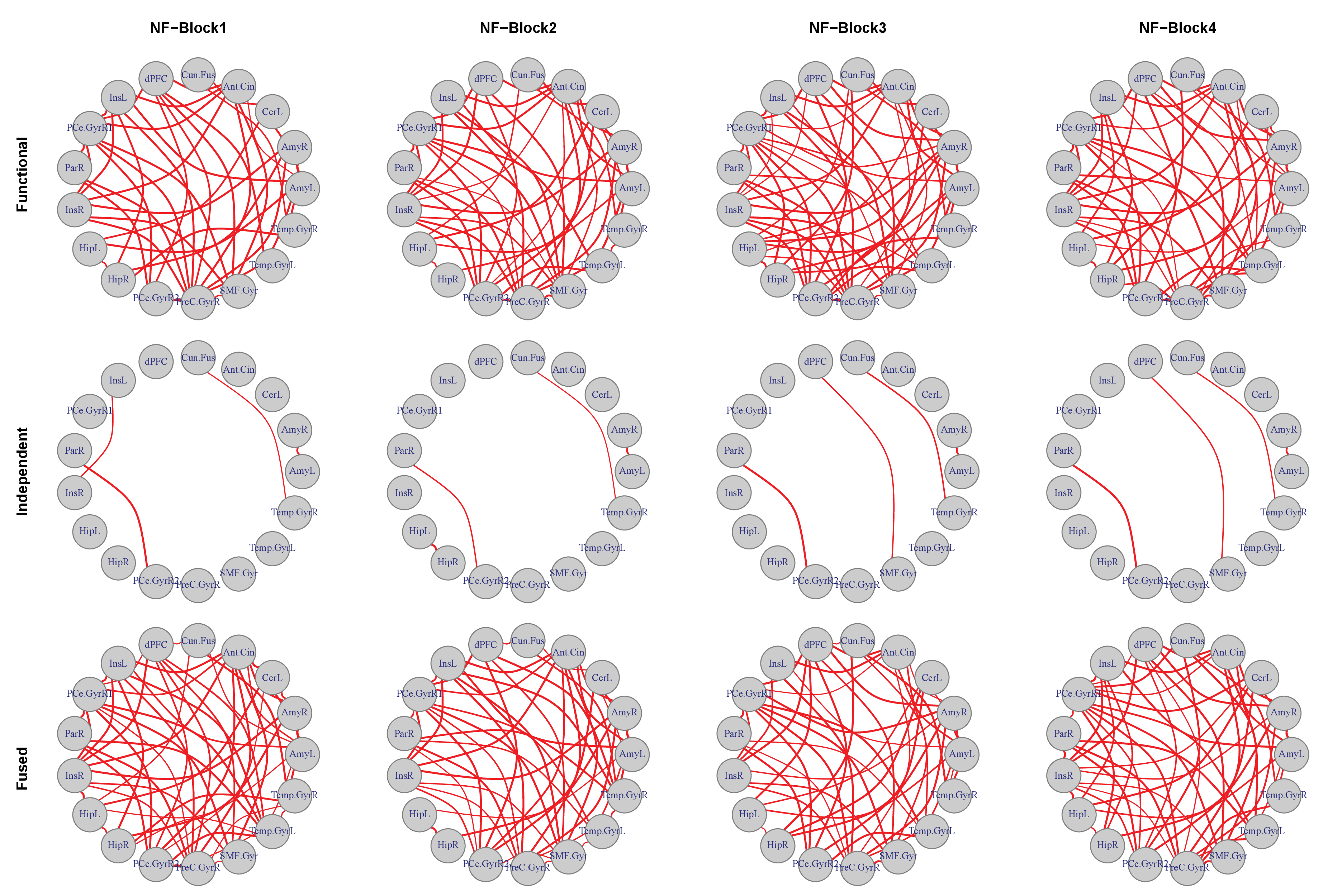}}
\caption{The average networks over the four blocks of neurofeedback training inferred by the Bayesian functional graphical model (upper panel), Bayesian independent graphical model (middle panel), and fused graphical lasso (lower panel) among the depressed adolescents.}
\label{fig:rtfMRI_network}
\end{figure}

Figure \ref{fig:rtfMRI_network} displays depressed adolescents' average FC networks over the four blocks of NF-task training blocks inferred by the Bayesian functional graphical model (upper panel), Bayesian independent graphical model (middle panel), and fused graphical lasso (lower panel). The movie showing the inferred FC network changes over the task fMRI is provided in the supplementary materials. We observe that the Bayesian independent graphical model failed to detect FCs between regions likely because it fails to learn from nearby locations by accounting for within-functional correlations over time. Both the Bayesian functional graphical model and fused graphical lasso identified many brain connections that are consistently present over time, for example, the FC between right-amygdala and left-hippocampus and that between left-amygdala and right-hippocampus. However, the inferred networks of our method show an obvious increase in connection density over time while those of the fused graphical lasso did not. The increased network density indicates that the NF-task training blocks were able to strengthen FCs between key brain regions. 

\begin{figure}
\centerline{\includegraphics[height=5.0in,width=4.0in]{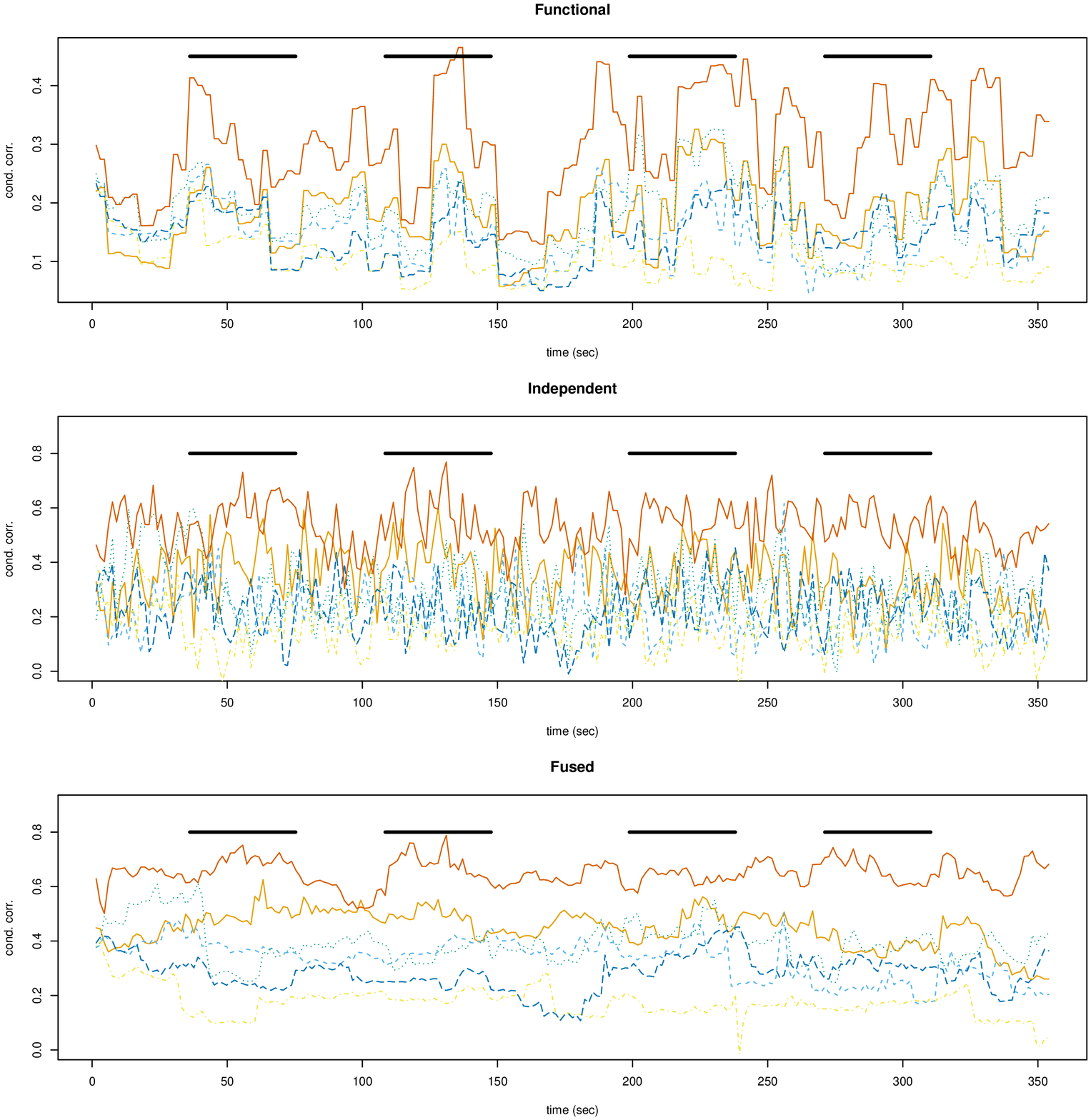}}
\caption{Plot of point estimates of conditional cross-correlations versus time obtained by the Bayesian functional graphical model (upper panel), Bayesian independent graphical model (middle panel), and fused graphical lasso (lower panel) among the depressed adolescents. The black solid segments at the top indicate the times of the four NF-task blocks during fMRI.}
\label{fig:rtfMRI_ccor}
\end{figure}

This observation is also confirmed by the plots of selected conditional cross-correlations over time as displayed in Figure \ref{fig:rtfMRI_ccor}. The figure plots the estimated conditional cross-correlations of five identified FCs versus time among depressed adolescents, obtained by the Bayesian functional graphical model (upper panel), Bayesian independent graphical model (middle panel), and fused graphical lasso (lower panel). The black solid segments at the top indicate the four NF-task blocks during fMRI. We observe that the time-varying conditional cross-correlations of the Bayesian functional graphical model show blocks of obviously elevated dependency that are approximately in align with the blocks of NF-task training. In addition, these blocks of elevated correlations increase in magnitude over time, indicating the NF-task training strengthened FCs between brain regions. However, the trends were not or only vaguely observed for the Bayesian independent graphical model and fused graphical lasso. 

We also applied the Bayesian functional graphical model to the fMRI data collected from healthy controls, which did not detect many connections likely due to the small sample size. The sparse network and conditional cross-correlation estimates as shown in Figures S5 and S6 in the supplementary materials indicate decreased FCs between brain regions over time during the NF-task fMRI.

\section{Discussion} \label{sec:diss}
In this article, we propose a Bayesian functional graphical model that obtains estimates of network structures that evolve smoothly over the functional domain. Our approach utilizing the basis transformation strategy and a variant of the Bayesian graphical lasso method for graphical modeling in dual basis space, which induces a N-Hypo mixture prior for adaptive shrinkage estimation of the conditional cross-covariance matrices in the data space. To our knowledge, we are the first to introduce the N-Hypo as a new normal-scale mixture distribution and characterize its shrinkage properties. We show that this mixture prior provides adaptive shrinkage that pools information both locally and globally, and that near-lossless basis representation with potentially a small number of basis functions is sufficient for functional graphical model estimation. The computational time of our method is linear in $K(\leq T)$ and quadratic in $p$, and can be greatly improved by parallel computing. Using 24 computing cores, it only takes about 6 and 20 minutes for 5000 iterations in the simulation ($p=10,T=128$) and fMRI data analysis ($p=17, T=235$) respectively.

We conducted a simulation study in which data were generated from non-stationary multivariate Gaussian processes in presence of local or global within-functional correlations. The results showed that our Bayesian functional graphical models using wavelet basis functions well approximate the true functional changes of the connection networks in the autocorrelated Gaussian process. Comparing to competitive methods that do not account for within-functional correlations or only uses local information, our method showed balanced performance in edge detecting with both high powers and low type 1 error rates for scenarios with either short- or long-range autocorrelations. The simulation studies also showed that our method is able to detect connections among variables across two different locations, which might shed some lights on the pattern of autocorrelation. The real data analysis of the NF-task fMRI for adolescent depression also showed that our Bayesian functional graphical model was able to detect the changes in FC networks across time by pooling information from both local and global functional domain. 

From the frequentist point of view, the functional graphical model we introduce in basis space is equivalent to applying an $L_1$ constraint to the conditional covariance between basis coefficients, i.e.$\sum_{j \neq \ell} |c^*_{k,j\ell}| < \alpha_k$ for some constraint $\alpha_k$. This leads to an upper bound on the $L_1$ norm of the conditional cross-covariance matrix in data space as 
\bse
\sum_{j \neq \ell}|C_{j\ell}(t,t)| &=& \sum_{j \neq \ell} \Bigg| \sum_k \phi^2_k(t)(s^*_{kj}s^*_{k\ell})^{-1}c^*_{k,j\ell} \Bigg| \\
& \leq & \sum_k \phi^2_k(t)(s^*_{kj}s^*_{k\ell})^{-1}\left(\sum_{j \neq \ell} |c^*_{k,j\ell}| \right) < \sum_k \phi^2_k(t)(s^*_{kj}s^*_{k\ell})^{-1}\alpha_k,
\ese
which implies an induced $L_1$ type of shrinkage in the data space by our method. However, we cannot use the frequentist graphical lasso for model estimation in the basis space, since the data-space conditional cross-covariance at $t$ is zero only if the two variables have zero conditional cross-correlations for all the basis functions that are supported at $t$ as indicated by Lemma 1. Therefore sparsity in the basis space does not ensure sparsity of graphical models in the data space. However, shrinkage in the basis space induced by our Bayesian method does lead to shrinkage of conditional cross-covariance matrix in the data space and sparse graphical estimations with appropriate posterior inference.

We implemented our method in R and have posted the codes in GitHub (\url{https://github.com/zhan4800/FunGraph}). While we use wavelets, the methodology is general for other basis functions such as functional principal components, Fourier bases, or splines depending on the applications. The method can also be used with other shrinkage priors which we expect to have similar adaptive shrinkage property as the lasso prior. Future directions of the work include implementing the functional graphical model in a functional regression framework that incorporate covariates and extending to discrete graphical models with general classes of distributions.

\bibliographystyle{rss} 
\bibliography{refs}

\section*{Supplement}
\startsupplement

\subsection*{Appendix A: Proof of Lemma 3}

We use moment generating function (MGF) to prove Lemma 3. We first look at the MGF of hypoexponential distributions. Suppose $\tau^2_1,\ldots, \tau^2_K$ independently follow exponential distributions $\mathrm{Exp}(\lambda_1), \ldots, \mathrm{Exp}(\lambda_K)$ respectively. By definition, $\tau^2=\sum^K_{k=1}\tau^2_k$ has a hypoexponential distribution $\mathrm{Hypo}(\lambda_1/2,\ldots,\lambda_K/2)$. The MGF of $\tau^2$ is then
\bse
M_{\tau^2}(t) = E\[e^{t \tau^2}\] = E\[e^{t \sum^K_{k=1}\tau^2_k}\] = \prod^K_{k=1}E\[e^{t\tau^2_k}\] = \prod^K_{k=1} M_{\tau^2_k}(t) = \prod^K_{k=1}\frac{\lambda_k}{\lambda_k-2t}.
\ese 
Let $X$ be a random variable following the normal-hypoexponential-scale-mixture distribution with parameters $\lambda_1,\ldots,\lambda_K$. The MGF of $X$ is
\bse
M_X(t) &=& \int e^{tx} f(x) dx \\
		&=& \int e^{tx} \int \mathcal{N}(x|0,\tau^2) \mathrm{Hypo}(\tau^2|\lambda_1,\ldots,\lambda_K) d\tau^2 \:dx \\
		&=& \int e^{tx} \int \mathcal{N}(x|0,\tau^2) dx \:\mathrm{Hypo}(\tau^2|\lambda_1,\ldots,\lambda_K) d\tau^2 \\
		&=& \int e^{\tau^2 t^2/2} \: \mathrm{Hypo}(\tau^2|\lambda_1,\ldots,\lambda_K) d\tau^2  \\
		&=& M_{\tau^2}(\frac{t^2}{2}) \\
		&=& \prod^K_{k=1}\frac{\lambda_k}{\lambda_k-t^2}.
\ese
Now let's look at the MGF of sum of independent Laplace variables. Suppose $Y_1,\ldots,Y_K$ each follows an independent Laplace distribution with respective parameters $\lambda_1,\ldots,\lambda_K$. Then $Y=\sum^K_{k=1}Y_k$ has a MGF
\bse
M_Y(t) = E\[e^{t Y}\] = E\[e^{t \sum^K_{k=1}Y_k}\] = \prod^K_{k=1}E\[e^{tY_k}\] = \prod^K_{k=1} M_{Y_k}(t) = \prod^K_{k=1}\frac{\lambda_k}{\lambda_k-t^2},
\ese 
which is equivalent to that of $X$. Lemma 3 is proved. 

\subsection*{Appendix B. Proof of Theorems 1\&2} 

We first look at $m_k(\bar{y})$, the predictive density under a normal-exponential prior with a scale parameter $\lambda_k/2$.
\bse
m_k(\bar{y}) &=& \int \mathcal{N}(\bar{y}|\mu,\frac{1}{n}) \: \mathcal{N}(\mu|0,\tau^2) \: \mathrm{Exp}(\tau^2|\frac{\lambda_k}{2}) \: d\mu d\tau^2 \\
	&=& \frac{\lambda_k}{2\sqrt{2\pi}} \int^\infty_0 (\frac{1}{n}+\tau^2)^{-1/2} \: \exp \left\{ -\frac{1}{2} (\frac{1}{n}+\tau^2)^{-1}\bar{y}^2 - \frac{1}{2} \lambda_k \tau^2 \right\} \:d\tau^2 
\ese
We transform $\tau^2$ to $z=(\frac{1}{n}+\tau^2)^{-1}$ and obtain
\bse
m_k(\bar{y}) &=& \frac{\lambda_k}{2\sqrt{2\pi}} \exp \(\frac{\lambda_k}{2n}\) \int^n_0 \underbrace{z^{-3/2} \exp \( -\frac{1}{2}\bar{y}z-\frac{1}{2}\frac{\lambda_k}{z}\)}_{\mathrm{Inv-Gauss}\(z|\frac{\sqrt{\lambda_k}}{|\bar{y}|},\lambda_k\) } \:dz \\
 			&=& \underbrace{\frac{\sqrt{\lambda_k}}{2} \exp\(\frac{\lambda_k}{2n}\)}_{c_k} \Big\{ \underbrace{\exp(-\bar{y}\sqrt{\lambda_k}) \Phi(\sqrt{n}\bar{y}-\sqrt{\lambda_k/n})}_{m_{k1}(\bar{y})} + \underbrace{\exp(\bar{y}\sqrt{\lambda_k}) \Phi(-\sqrt{n}\bar{y}-\sqrt{\lambda_k/n})}_{m_{k2}(\bar{y})} \Big\}
\ese
where $\Phi(\cdot)$ indicates the cumulative density function (cdf) of a standard normal distribution. 

Now we obtain the derivative of $m_k(\bar{y})$.
\bse
m'_k(\bar{y}) = \frac{dm_k(\bar{y})}{d\bar{y}} =& c_k &\left\{ \frac{dm_{k1}(\bar{y})}{d\bar{y}} + \frac{dm_{k2}(\bar{y})}{d\bar{y}} \right\} \\
	=& c_k & \Big\{ \underbrace{-\sqrt{\lambda_k} \exp(-\sqrt{\lambda_k}\bar{y}) \Phi\(\sqrt{n}\bar{y}-\sqrt{\lambda_k/n}\)}_{m'_{k11}(\bar{y})} \\
	&& \underbrace{ + \sqrt{n} \exp(-\sqrt{\lambda_k}\bar{y}) \phi\(\sqrt{n}\bar{y}-\sqrt{\lambda_k/n}\)}_{m'_{k12}(\bar{y})} \\
	&& \underbrace{ \sqrt{\lambda_k} \exp(\sqrt{\lambda_k}\bar{y}) \Phi\(-\sqrt{n}\bar{y}-\sqrt{\lambda_k/n}\)}_{m'_{k21}(\bar{y})} \\
	&& \underbrace{ -\sqrt{n} \exp(\sqrt{\lambda_k}\bar{y}) \phi\(-\sqrt{n}\bar{y}-\sqrt{\lambda_k/n}\)}_{m'_{k22}(\bar{y})}	\Big\}
\ese

We then have
\bse
\lim_{\bar{y} \rightarrow \infty} S_k(\bar{y}) &=& \lim_{\bar{y} \rightarrow \infty}-\frac{d}{d\bar{y}}\log m_k(\bar{y}) = \lim_{\bar{y} \rightarrow \infty} -\frac{m'_k(\bar{y})}{m_k(\bar{y})} \\
	&=& -\lim_{\bar{y} \rightarrow \infty} \frac{m'_{k11}(\bar{y})+m'_{k12}(\bar{y})++m'_{k21}(\bar{y})+m'_{k22}(\bar{y})}{m_{k1}(\bar{y})+m_{k2}(\bar{y})} .
\ese
It's easy to show that all the six terms, $m'_{k11}(\bar{y}), m'_{k12}(\bar{y}), m'_{k21}(\bar{y}), m'_{k2}(\bar{y}), m_{k1}(\bar{y})$, and $m_{k2}(\bar{y})$, converges to 0 as $\bar{y} \rightarrow 0$, and 
\bse
m_{k2}(\bar{y})=o(m_{k1}(\bar{y})), && m'_{k12}(\bar{y}) = o(m_{k1}(\bar{y})),  \\
m'_{k21}(\bar{y})=o(m_{k1}(\bar{y})), &&m'_{k22}(\bar{y})=o(m_{k1}(\bar{y})).
\ese
Dividing each term by $m_{k1}(\bar{y})$, we have
\bse
\lim_{\bar{y} \rightarrow \infty} S_k(\bar{y}) = - \lim_{\bar{y} \rightarrow \infty} \frac{m'_{k11}(\bar{y})}{m_{k1}(\bar{y})} = -\sqrt{\lambda_k}.
\ese
Theorem 1 proved.

Now we look at $m(\bar{y})$, the predictive density under a normal-hypoexponential prior with the parameter set $(\lambda_1/2,\ldots,\lambda_K/2)$.
\bse
m(\bar{y}) &=& \int \mathcal{N}(\bar{y}|\mu,\frac{1}{n})\mathcal{N}(\mu|0,\tau^2) \: \mathrm{Hypo}(\tau^2|\lambda_1,\ldots,\lambda_K) \: d\mu d\tau^2 \\
	&=& \frac{1}{\sqrt{2\pi}} \int^\infty_0 (\frac{1}{n}+\tau^2)^{-1/2} \exp \left\{ -\frac{1}{2}(\frac{1}{n}+\tau^2)^{-1} \bar{y}^2 \right\} \sum^K_{k=1}P_k \frac{\lambda_k}{2} \exp \left\{ -\frac{\lambda_k}{2}\tau^2 \right\} \:d\tau^2 \\
	&=& \sum^K_{k=1} P_k \frac{\lambda_k}{2\sqrt{2\pi}} \int^\infty_0 (\frac{1}{n}+\tau^2)^{-1/2} \: \exp \left\{ -\frac{1}{2} (\frac{1}{n}+\tau^2)^{-1}\bar{y}^2 - \frac{1}{2} \lambda_k \tau^2 \right\} \:d\tau^2 \\
	&=& \sum^K_{k=1} P_k m_k(\bar{y}),
\ese
which is a linear combination of $m_k(\bar{y})$. Therefore,
\bse
S(\bar{y}) &=& -\frac{d}{d\bar{y}}\log m(\bar{y}) = \frac{dm(\bar{y})/d\bar{y}}{m(\bar{y})} \\
		&=& -\frac{\sum^K_{k=1}P_k \: m'_k(\bar{y})}{\sum^K_{k=1}P_k \:m_k(\bar{y})} \\
		&=& -\frac{\sum^K_{k=1}P_k \{m'_{k11}(\bar{y})+m'_{k12}(\bar{y})++m'_{k21}(\bar{y})+m'_{k22}(\bar{y}) \} }{\sum^K_{k=1}P_k \{ m_{k1}(\bar{y})+m_{k2}(\bar{y}) \} } ,
\ese
where for each $k=1,\ldots,K$, $m'_{k11}(\bar{y}), m'_{k12}(\bar{y}), m'_{k21}(\bar{y}), m'_{k2}(\bar{y}), m_{k1}(\bar{y})$, and $m_{k2}(\bar{y})$, converges to 0 as $\bar{y} \rightarrow 0$.  

Let $\kappa=\arg\!\min \{\lambda_k:k=1,\ldots,K \}$. That is, $\lambda_\kappa < \lambda_k  \text{ for all } k\neq \kappa$. Thus for all $k \neq \kappa$, we have
\bse
\lim_{\bar{y}\rightarrow \infty} \frac{m_{k1}(\bar{y})}{m_{\kappa1}(\bar{y})} = \lim_{\bar{y}\rightarrow \infty}  \(\frac{\lambda_k}{\lambda_\kappa}\)^{1/2} \exp \left\{-\(\lambda_k^{1/2}-\lambda_\kappa^{1/2}\)\bar{y} \right\} \frac{\Phi\(\sqrt{n}\bar{y}-\sqrt{\lambda_k/n}\)}{\Phi\(\sqrt{n}\bar{y}-\sqrt{\lambda_\kappa/n}\)} =0,
\ese
since $\lambda_k>\lambda_\kappa$. Hence $m_{k1}(\bar{y})=o(m_{\kappa1}(\bar{y}))$, and $m'_{k11}(\bar{y})=o(m_{\kappa1}(\bar{y}))$ likewise. In addition, because
\bse
m_{k2}(\bar{y})=o(m_{k1}(\bar{y})), && m'_{k12}(\bar{y}) = o(m_{k1}(\bar{y})),  \\
m'_{k21}(\bar{y})=o(m_{k1}(\bar{y})), &&m'_{k22}(\bar{y})=o(m_{k1}(\bar{y})),
\ese
and $m_{k1}(\bar{y})=o(m_{\kappa1}(\bar{y}))$, we have
\bse
m_{k2}(\bar{y})=o(m_{\kappa 1}(\bar{y})), && m'_{k12}(\bar{y}) = o(m_{\kappa 1}(\bar{y})),  \\
m'_{k21}(\bar{y})=o(m_{\kappa 1}(\bar{y})), &&m'_{k22}(\bar{y})=o(m_{\kappa 1}(\bar{y})).
\ese
We then have
\bse
\lim_{\bar{y}\rightarrow \infty} S(\bar{y}) = - \lim_{\bar{y}\rightarrow \infty} \frac{P_\kappa m'_{\kappa 11}(\bar{y})}{P_\kappa m_{\kappa 1}(\bar{y})} = -\sqrt{\lambda_\kappa}.
\ese
Theorem 2 proved.

\newpage
\subsection*{Supplementary figures and tables for simulation}

\begin{figure}[h]
\centering
\centerline{\includegraphics[width=4in]{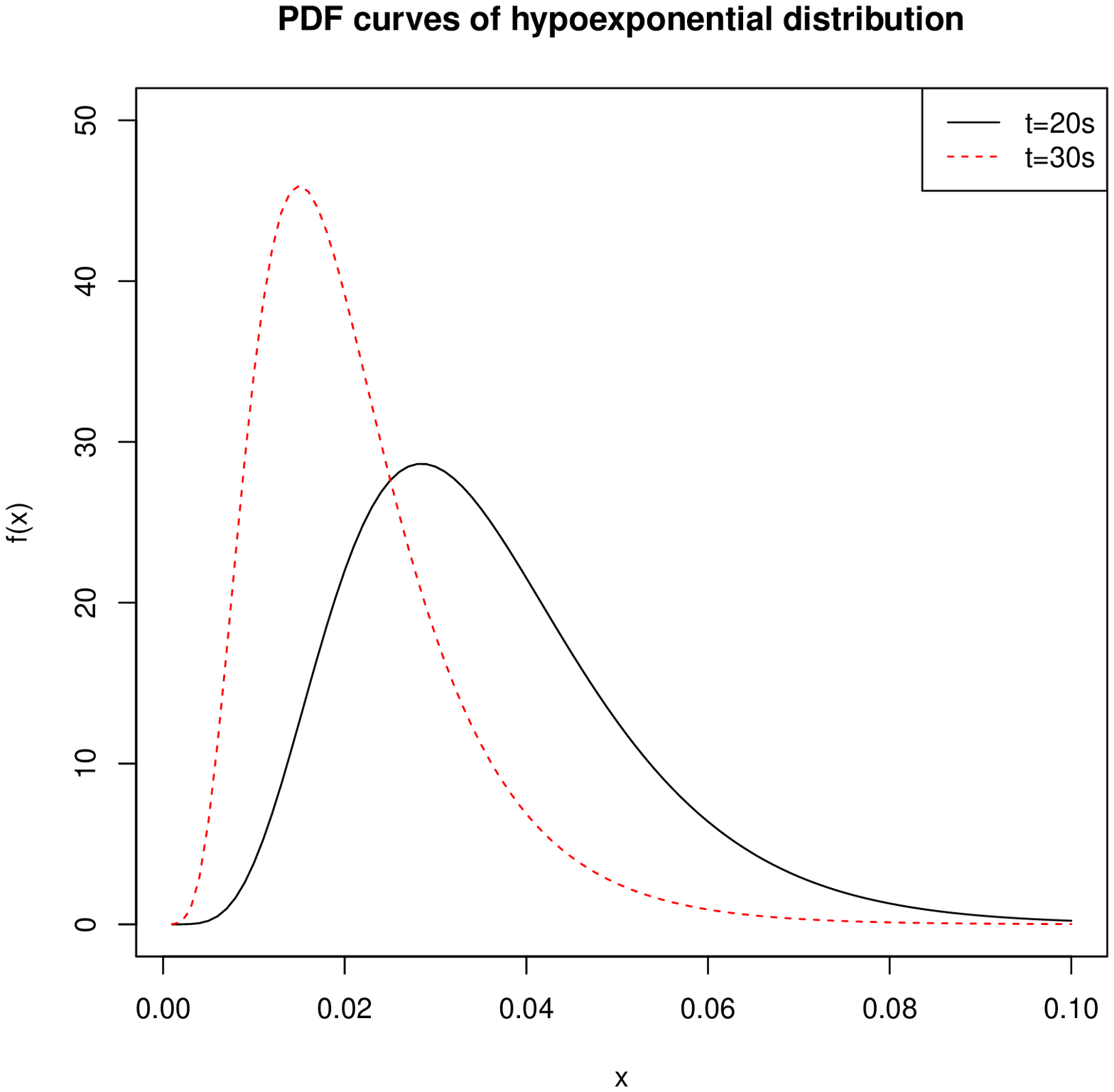}}
\caption{The density curves of the hypoexponential distribution at two time points in the NF-task fMRI data analysis.}
\label{fig:hypo_density}
\end{figure}

\begin{figure}[h]
\centering
\centerline{\includegraphics[width=6.5in]{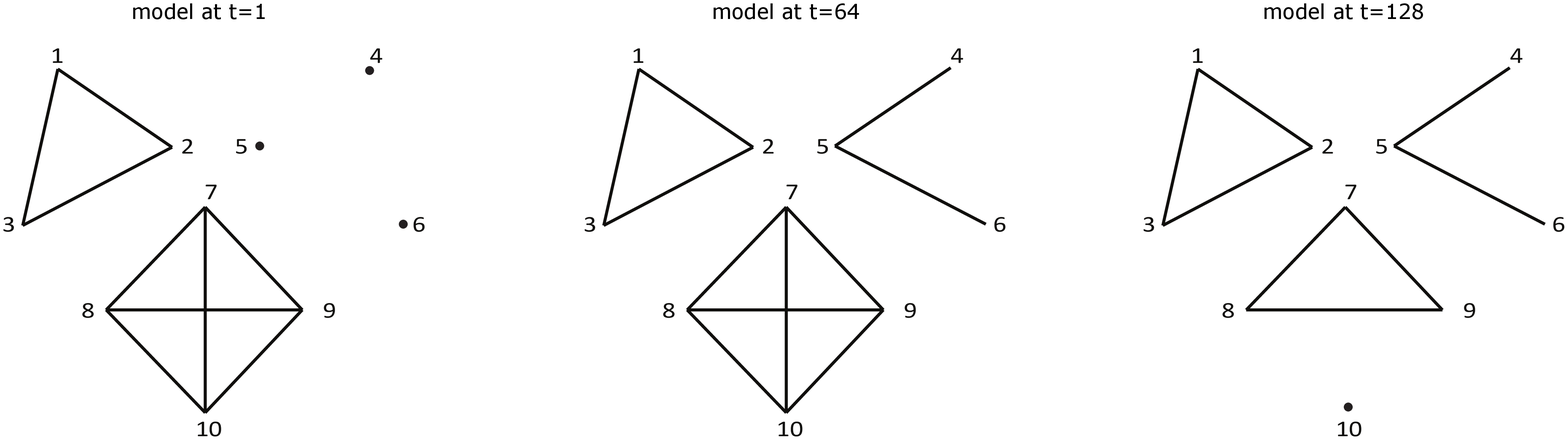}}
\caption{The graphical models at three different time points, $t=1$, $64$, and $128$, for dynamic model 1 in the simulation study.}
\label{fig:simu_graphs_illu}
\end{figure}

\newpage

\begin{figure}[h]
\centering
\centerline{\includegraphics[width=6.5in]{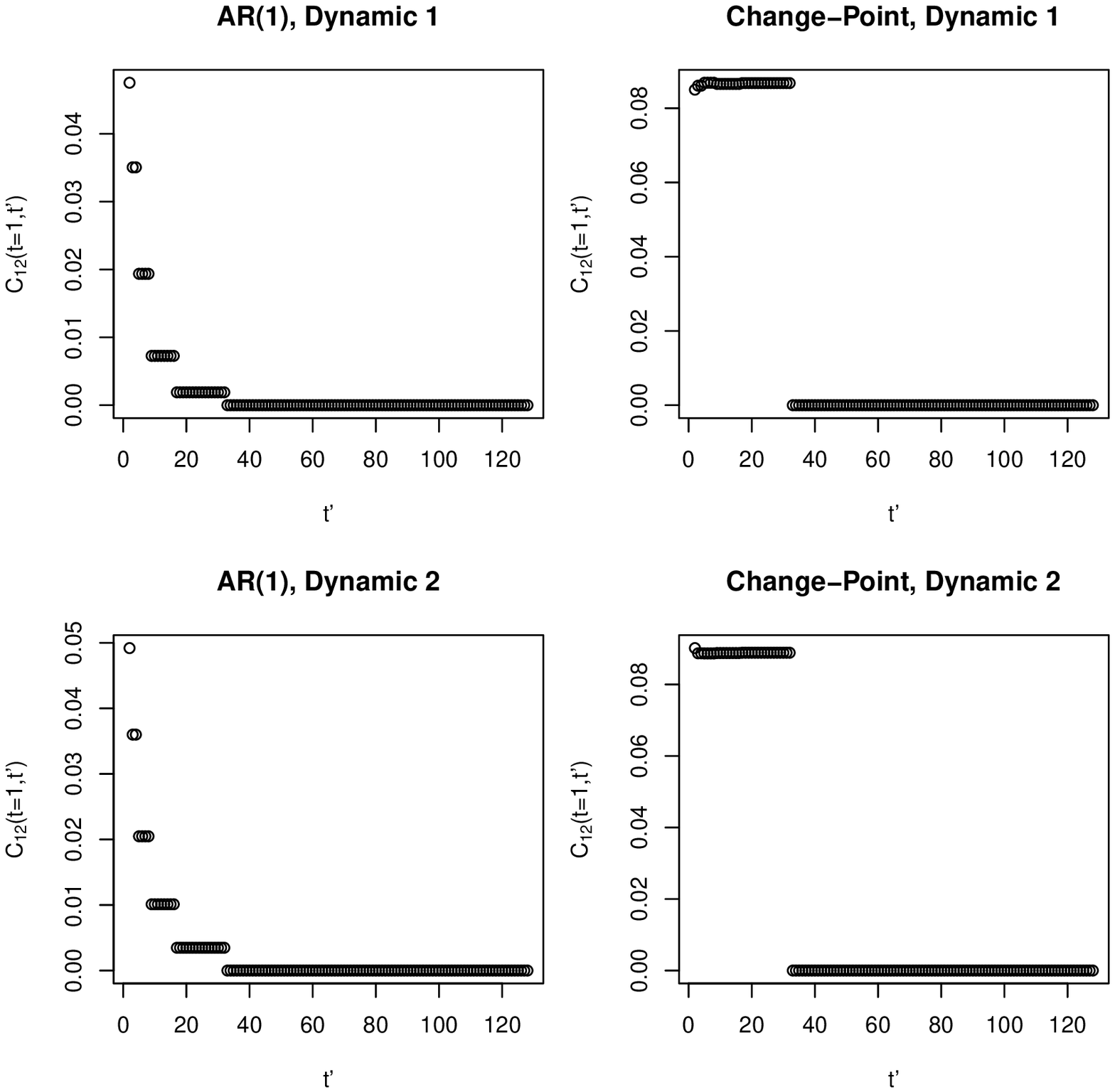}}
\caption{Estimated lagged conditional cross-covariance between Nodes 1\&2, $C_{12}(t,t^\prime)$, for $t=1$ and $t^\prime=2,\ldots,T$, by the Bayesian functional graphical model, averaged across 100 simulated datasets.}
\label{fig:simu_EROC}
\end{figure}

\newpage

\begin{table}[h]
\begin{center}
\begin{tabular}{|ll|}
\hline
{ROI} & {Hemisphere}\\
\hline
{Amygdala} & {Left} \\
{Amygdala} & {Right} \\
{Cerrebelum} & {Left} \\
{Anterior Cingulate Cortex} & {Left and Right} \\
{Cuneus, Fusiform} & {Right} \\
{dPFC} & {Left} \\
{Hippocampus} & {Left} \\
{Hippocampus} & {Right} \\
{Inferior Parietal Lobe} & {Right} \\
{Insula, Inferior Frontal Gyrus} & {Left} \\
{Insula, Inferior Frontal Gyrus, Putamen} & {Right} \\
{Postcentral Gyrus} & {Right} \\
{Precentral, Middle, and Inferior Frontal Gyrus} & {Right} \\
{Superior and Middle Frontal Gyrus} & {Right} \\
{Superior, Middle and Inferior Temporal Gyrus} & {Left} \\
{Superior, Middle and Inferior Temporal Gyrus} & {Right}  \\
\hline
\end{tabular}
\caption{ROIs in NF-task fMRI data analysis.} \label{tab:real_roi}
\end{center}
\end{table}

\newpage

\begin{figure}[h]
\centering
\centerline{\includegraphics[width=6.5in]{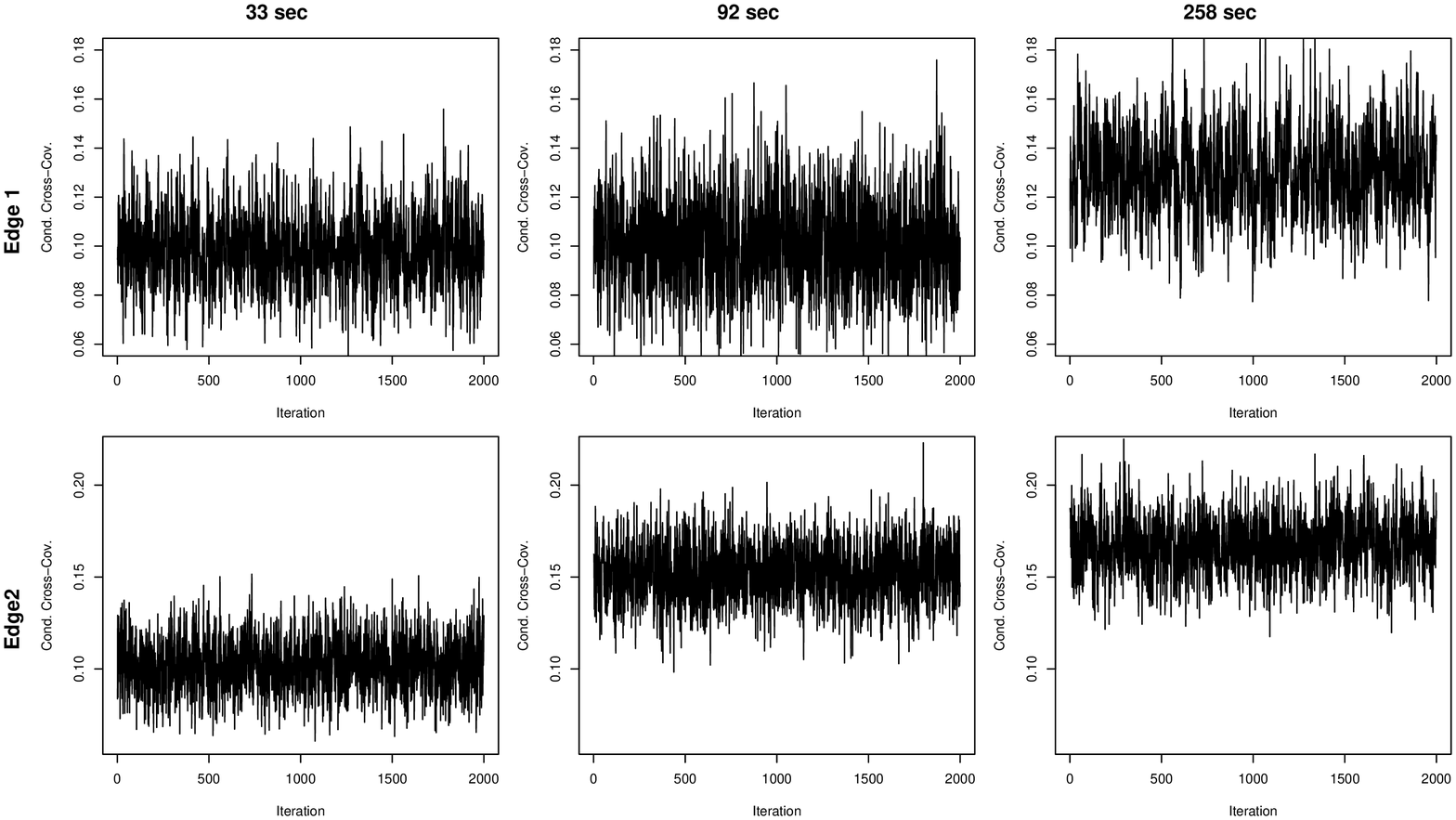}}
\caption{The traceplots of conditional cross-covariances of two identified edges at three time points in the NF-task fMRI analysis.}
\label{fig:real_traceplots}
\end{figure}

\newpage

\begin{figure}[h]
\centering
\centerline{\includegraphics[width=6.5in]{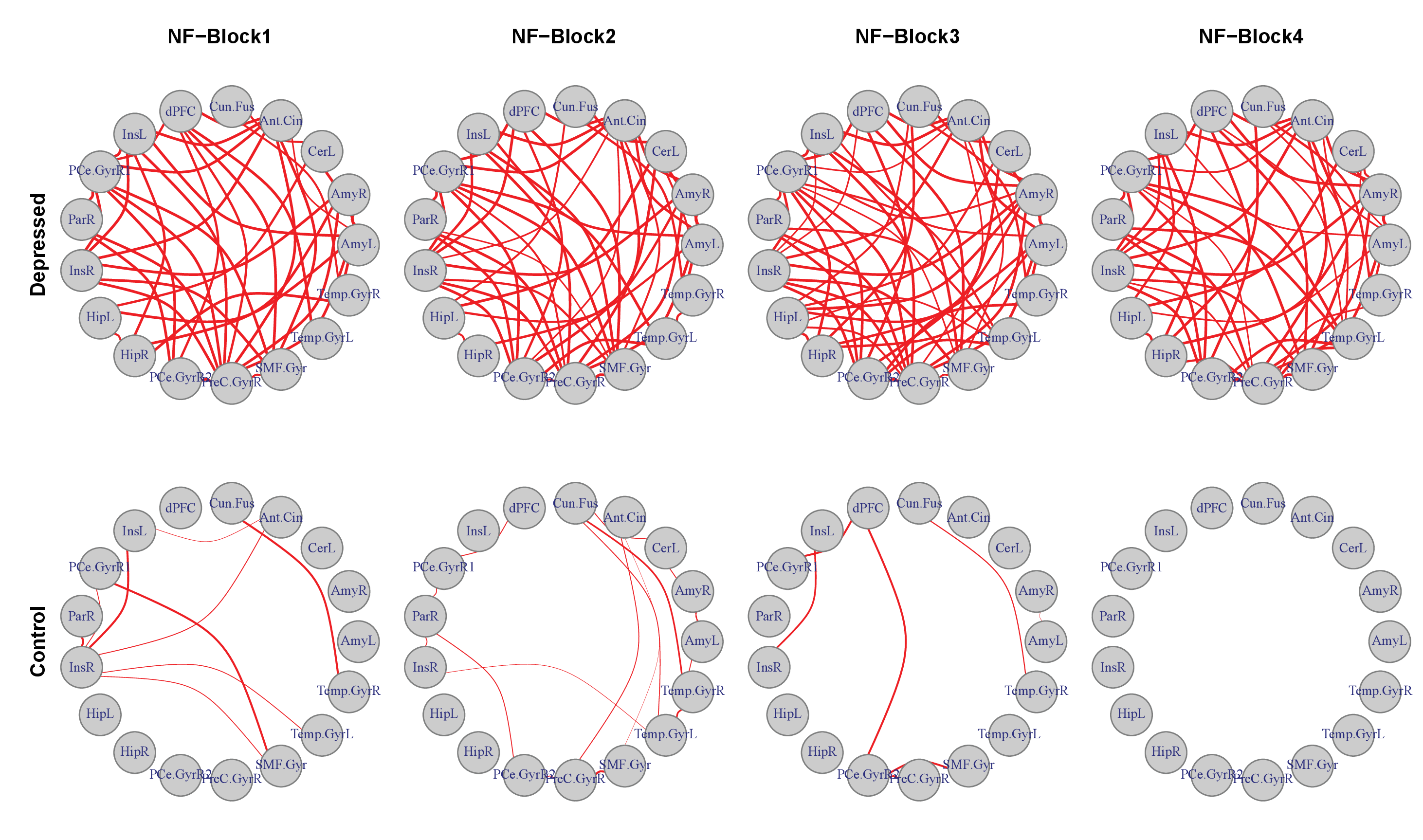}}
\caption{The average networks over the four blocks of NF training inferred by the Bayesian functional graphical mode for the depressed adolescents (upper panel) versus healthy controls (lower panel).}
\label{fig:real_network_DPvsCT}
\end{figure}

\newpage

\begin{figure}
\centerline{\includegraphics[height=5.0in,width=4.0in]{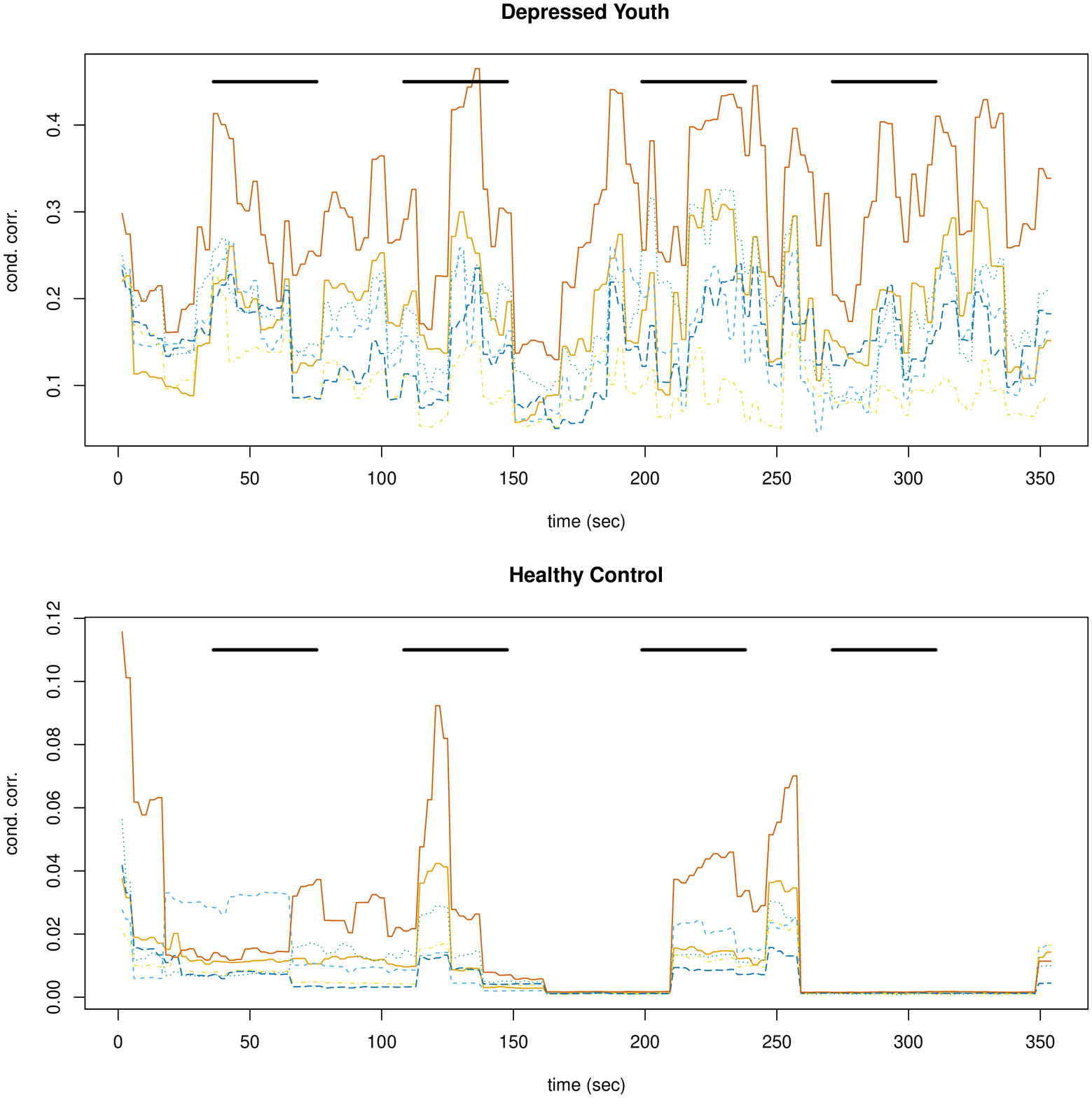}}
\caption{Plot of point estimates of conditional cross-correlations versus time obtained by the Bayesian functional graphical model for the depressed adolescents (upper panel) versus healthy controls (lower panel). The black solid segments at the top indicate the times of the four NF-task blocks during fMRI.}
\label{fig:rtfMRI_ccor_DPvsCT}
\end{figure}

\end{document}